\documentclass[useAMS, usenatbib]{mn2e}

\usepackage{graphicx}
\usepackage{epsfig}
\usepackage{amsbsy}
\usepackage{amsmath}  
\usepackage{amssymb}
\usepackage{subfigure}
\usepackage{array}

\makeatletter
  \renewcommand{\@thesubfigure}{\,\,\,\,\,\,\,\,\,\,(\alph{subfigure})}
  \renewcommand{\p@subfigure}{\thefigure}
\makeatother

\newcommand{\bayesys}{\texttt{BayeSys}}           

\def\reff@jnl#1{{\rm#1\/}}

\def\aj{\reff@jnl{AJ}}                  
\def\araa{\reff@jnl{ARA\&A}}            
\def\apj{\reff@jnl{ApJ}}                        
\def\apjl{\reff@jnl{ApJ}}               
\def\apjs{\reff@jnl{ApJS}}              
\def\aap{\reff@jnl{A\&A}}               
\def\mnras{\reff@jnl{MNRAS}}            

\begin{document}

\title[]{Very Small Array observations of the Sunyaev--Zel'dovich
effect in nearby galaxy clusters}

\author[]{Katy Lancaster$^{1,7}$,
  Ricardo Genova-Santos$^2$, Nelson Falc{\`o}n$^{2,6}$, Keith
  Grainge$^1$,
\newauthor Carlos Guti{\`e}rrez$^2$, R{\"u}diger Kneissl$^1$,
  Phil Marshall$^4$, Guy Pooley$^1$, 
\newauthor Rafael Rebolo$^{2,5}$, Jose-Alberto Rubi{\~ n}o-Martin$^2$,
  Richard D. E. Saunders$^1$, 
\newauthor Elizabeth Waldram$^1$, Robert A. Watson $^3$.\\
$^1$Astrophysics Group, Cavendish Laboratory, University of Cambridge,
Madingley Road, Cambridge CB3 0HE\\ $^2$Instituto de Astrofis\'{i}ca
de Canarias, 38200 La Laguna, Tenerife, Canary Islands, Spain\\
$^3$Jodrell Bank Observatory, University of Manchester, Macclesfield,
Cheshire SK11 9DL \\ $^4$Kavli Institute for Particle Astrophysics and
Cosmology, Stanford University PO Box 20450, MS29, Stanford, CA 94309,
USA\\ $^5$Consejo Superior de Investigaciones Cientificas, Spain\\
$^6$ Dpt. de Fisica, FACYT, Universidad de Carabobo, Venezuela\\ $^7$
Present address: Physics Department, University of Bristol, Tyndall
Avenue, Bristol BS8 1TL}
\date{Received **insert**; Accepted **insert**}

\pagerange{\pageref{firstpage}--\pageref{lastpage}} 
\pubyear{}

\maketitle
\label{firstpage}

\begin{abstract}
We present Very Small Array (VSA) observations (centred on $\approx$
34\,GHz) on scales $\approx$ 20 arcmin towards a complete,
X-ray--flux--limited sample of seven clusters at redshift
$z<0.1$. Four of the clusters have significant Sunyaev-Zel'dovich (SZ)
detections in the presence of CMB primordial anisotropy. For all
seven, we use a Bayesian Markov-Chain-Monte-Carlo (MCMC) method for
inference from the VSA data, with X-ray priors on cluster positions
and temperatures, and radio priors on sources. In this context, the CMB
primordial fluctuations are an additional source of Gaussian noise,
and are included in the model as a non--diagonal covariance matrix
derived from the known angular power spectrum. In addition, we make
assumptions of $\beta$--model gas distributions and of hydrostatic
equilibrium, to evaluate probability densities for the gas mass
($M_{\rm{gas}}$) and total mass ($M_{\rm{r}}$) out to $r_{200}$, the
radius at which the average density enclosed is 200 times the critical
density at the redshift of the cluster.  This is further than has been
done before and close to the classical value for a collapsed
cluster. Our combined estimate of the gas fraction
$(f_{\rm{gas}}=M_{\rm{gas}}/M_{\rm{r}})$ is
$0.08^{+0.06}_{-0.04}h^{-1}$. The random errors are poor (note however
that the errors are higher than would have been obtained with the
usual chi-squared method on the same data) but the control of bias is
good. We have described the MCMC analysis method specifically in terms
of SZ but hope the description will be of more general use. We find
that the effects of primordial CMB contamination tend to be similar in
the estimates of both $M_{\rm{gas}}$ and $M_{\rm{r}}$ over the
narrow range of angular scales we are dealing with, so that there is
little effect of primordials on $f_{\rm{gas}}$ determination. Using
our $M_{\rm{r}}$ estimates we find a normalisation of the mass --
temperature relation based on the profiles from the VSA cluster
pressure maps that is in good agreement with recent $M-T$
determinations from X-ray cluster measurements.

\end{abstract}

\begin{keywords}
 cosmology: observations -- cosmic microwave background --  galaxies:
 clusters: individual (Coma, A1795, A399, A401, A478, A2142, A2244) --
 X-rays:galaxies:clusters
\end{keywords}


\section{INTRODUCTION}
\label{sec:introduction}

Galaxy clusters have long been thought to provide a faithful sample of
cosmic baryonic matter (see e.g. \cite{COS/Whi++93}, \cite{Evrard}).
One quantity often calculated and assessed in such work is the gas
fraction $f_{\rm{gas}}$, which is defined as the (baryonic) gas mass
over the total (baryonic plus dark matter) mass of the cluster.  We
here present Sunyaev Zel'dovich (SZ) (\cite{SZ}, see also e.g.
\cite{Birkinshaw1999}, \cite{Carlstrom_review}) observations of a sample of
clusters, from which we infer $f_{\rm{gas}}$. Our random errors are
high but the sample is complete, the redshifts deliberately low, and
we are able to estimate $f_{\rm{gas}}$ out to radii at which the
overdensity of the enclosed region is close to the classical value
of 178 for a collapsed object (see e.g. \cite{Peacock}). First we
review some of the existing $f_{\rm{gas}}$ measurements.

A popular route in investigating cosmic baryonic matter is the detailed
study of the X-ray emission from cluster gas. For example, in an
investigation based on \emph{ROSAT} PSPC data (\cite{fabian_fgas}), a
sample of 36 clusters of redshift $0.05 \le z \le 0.44$ was used to
measure $f_{\rm{gas}}$.  Assumptions of isothermality and hydrostatic
equilibrium were required.  The resulting $f_{\rm{gas}}$ distribution
(within $r_{500}$, that is, where the mean density inside this radius
is 500 times the critical density at the redshifts of the clusters)
was centred on a value $f_{\rm{gas}}(r_{500})=0.168h_{50}^{-1.5}$.
Values for individual clusters were found to vary between 0.101 and
0.245. \cite{Mohr} also analysed PSPC data on 45 X-ray selected
clusters, finding a mean $f_{\rm{gas}}(r_{500})$ of
$0.212h_{50}^{-1.5}$ in a subsample of 27 clusters hotter than 5 keV.
\cite{IOA1}, following a similar route (supplemented by gravitational
lensing information on the total mass) with \emph{Chandra} imaging
spectrometer data find, for a set of six clusters with $0.103 \le z
\le 0.461$, a mean $f_{\rm{gas}}$ within $r_{2500}$ of $0.113 \pm
0.005h_{70}^{-1.5}$ for a $\Lambda$--CDM model, a very precise
determination with very similar values for each cluster.  \cite{IOA2},
with additional data, investigated the observed change of
$f_{\mathrm{gas}}$ with cosmology.

Studies making use of the SZ effect have potential advantages for gas
and gravitational potential measurements (where the potential is
obtained via calculation of the total mass).  The X-ray signal is
proportional to $n_{\rm{e}}^2$ (where $n_{\rm{e}}$ is electron
density), while the SZ signal is proportional to $n_{\rm{e}}$.  This
means that SZ is less biased to concentration and can constrain
clumping. Although X-ray telescopes achieve excellent signal to noise,
they are restricted to observing the denser, inner regions of a cluster
(e.g out to $r_{2500}$).  With SZ it is possible to measure
$n_{\rm{e}}(r)$ over a larger range of r (e.g. close to the virial radius) as
less dynamic range is required.

\cite{Myers_obs} used the OVRO 5.5m telescope to observe the SZ effect
in 3 clusters at 32\,GHz.  With the addition of the Coma cluster
(observed by \cite{Herbig}), they obtain a gas fraction of
$f_{\rm{gas}}=0.061\pm{0.011}h^{-1}_{100}$ This sample of objects lies
in the redshift range $0.023 \leq z \leq 0.0899$, and includes three
clusters which we also present here.  (\cite{Mason_obs} extend the
sample to seven clusters, incorporating a further two discussed in
this paper.  The data were used to calculate $H_0$.)

\cite{grego} used the OVRO and BIMA arrays to make SZ observations of
galaxy clusters at 30\,GHz.  The data were used to infer the gas mass
and total mass, thus constraining $f_{\rm{g}}$ (within $r_{500}$) in 18
X-ray selected clusters in the redshift range $0.171\leq z \leq 0.826$.  The
mean value obtained for the full sample was
$f_{\rm{gas}}=0.081^{+0.009}_{-0.011}h^{-1}_{100}$.  In addition, a
`fair' subsample is defined as the five most X-ray luminous clusters
in the EMSS sample.  These objects have redshift $0.328\leq z \leq
0.826$, and together give a mean gas fraction
$f_{\rm{gas}}=0.089^{+0.018}_{-0.019}h^{-1}_{100}$.
 
One of the aims of the VSA project (\cite{paper1}, \cite{paper2},
\cite{paper3}, \cite{paper4}, \cite{paper5}, \cite{anze} \cite{clive},
\cite{rafa}) has been to image nearby, massive clusters in SZ. The VSA
baselines at $\approx34\rm{\,GHz}$ couple well to the angular scales
of such clusters. Here we describe SZ observations and
cluster--parameter inferences of an X-ray selected, complete sample of
seven clusters, with redshift $0.023 \leq z \leq 0.098$ and median
0.075. The age of the Universe at z = 0.075 is 1.7 times its age at $z =
0.55$.  The importance of low--z work is illustrated by the following
two points:

\begin{itemize}
\item {The low redshifts of the clusters mean that they have particularly
good X-ray data, and one can be reasonably confident that bright X-ray
selected complete samples are in fact complete.}
\item {Since clusters grow under gravity, then on average low redshift
clusters should be more evolved than those at higher
redshift. Comparison of, for example, $f_{\rm{gas}}$ in low- and
high-$z$ samples is important. (Of course, we do not know how big the
samples have to be to encompass meaningful averages).}
\end{itemize}

One immediate difficulty on these angular scales is contamination by
CMB primordial anisotropy. At the start of this VSA observational
programme, it was evident that we needed an analysis method that would
apply the inference process correctly and would properly cope with
error distributions in low signal--to--noise situations. There is the
additional difficulty of dealing with (potentially variable) radio
sources at 34\,GHz.  This could be especially problematic where
sources are in the clusters themselves rather than in the background:
the low redshifts of the clusters imply such sources may be very
bright.  Accounting for these effects correctly necessitates the
exploration of the posterior probabilities of the parameters of a
$\beta$--model for the gas distribution given the VSA visibilities,
receiver noise, the CMB and radio sources.  The method must also
incorporate prior knowledge on e.g. the cluster positions from X-rays,
and on source fluxes in a way which can cope with variability. We
assume isothermality, and that the clusters are well described by
hydrostatic equilibrium.  We use a Markov Chain Monte Carlo (MCMC)
sampler (\texttt{BayeSys}) for an acceptable combination of speed and
accuracy.

In section 2 we briefly describe the relevant features of the VSA.  In
section 3 we present the sample, outline the data reduction pipeline
and describe our strategy for dealing with radio sources.  In section
4 we present our results, and attempt to describe the Bayesian
analysis method in non-specialist terms.  We make concluding comments
in section 5.  


\section{The Very Small Array}

The VSA is a 14--element interferometric telescope situated at the
Observatorio del Teide, Tenerife.  The observing frequency is tunable
in the 26--36\,GHz range, with a bandwidth of 1.5\,GHz; at these
frequencies observations should be relatively free from contamination
by Galactic foregrounds for fields at high Galactic latitude.  The 14
antennas are identical. They rotate independently and are mounted on a
tilting table thus allowing tracking in two dimensions.  The table is
surrounded by an aluminium shield to prevent groundspill.

The telescope was designed to operate in two configurations: Compact
(see e.g. \cite{paper1} for technical details) and Extended (see
\cite{paper5}).  All data in this paper were taken using the extended
configuration.  The Extended Array has 322--mm diameter illuminated
apertures, resulting in a primary beam of 2.0$^{\circ}$ FWHM when
operating at 34 \,GHz.  The horn arrangement on the table allows for a
range of baselines between approximately 40\,cm and 3\,m.  The
telescope is sensitive to angular sizes in the range $0.25^{\circ} <
\theta < 1.2^{\circ}$, and is ideal for observing low redshift
clusters.

Radio sources are a problem in all cm--wave CMB observations at all
but the lowest angular resolutions, and SZ is no exception.  The VSA
design includes a dedicated source--subtraction telescope.  This
comprises two 3.7\,m dishes located next to the main array and used as
an interferometer with a 9\,m baseline, giving 4 arcmin resolution and a
9 arcmin field of view.  The source--subtractor does not resolve any of
the sources which we observe, but resolves out the CMB fluctuations.


\section{Observations}

\subsection{Galaxy Clusters}

The VSA targets were selected from the Northern ROSAT All-Sky Survey
(\cite{NORAS}, NORAS hereafter) as the seven most X-ray luminous
objects at redshift $<0.1$. The clusters have rest--frame X-ray
luminosity $> 5 \times 10^{37}$W in the 0.1--2.4\,keV
energy band. Additionally, only clusters observable from Tenerife and
Cambridge were considered.  This imposed declination limits of
$10^{\circ} < \delta < 60^{\circ}$.  The upper limit is set by the
latitude and configuration of the VSA main array.  The lower limit is
set by the need for the use of the Ryle Telescope (RT) as part of the
source--subtraction strategy (see section \ref{sec:sources}).  Note
that we have \emph{not} applied any criteria concerning fluxes of
contaminant radio sources.  This is unlike the VSA primordial work,
and indeed the SZ work of the RT and OVRO/BIMA.

Pointing centres for the seven fields were defined based on the X-ray
positions of the clusters as published in NORAS.  Data for each target
were obtained in a series of short observations made during the period
October 2001--August 2003. Repeat observations were required in
several cases due to uncharacteristically persistent bad weather. The
sample is summarised in Table \ref{tab:clusters}, along with published
redshifts, temperatures used in our analysis, X-ray luminosities and
total integration times of the VSA observations.  The clusters A401
and A399 are only separated by around a degree, so were observed in a
single pointing centred on A401.

\begin{table*}
\centering

\caption{The VSA cluster sample: Cluster coordinates (\protect\cite{NORAS}),
redshift (\protect\cite{redshifts}), electron temperature (\protect\cite{Markevitch1998}),
except Coma, \protect\cite{hughes_coma}, X-ray luminosity
(\protect\cite{NORAS}), integration time, map rms (outside the primary
beam).}

\begin{tabular}{lcccccccc}
    \hline

Cluster	&RA	    &Dec	&$z$	&$T_{e}$	&$L_{X}$ 		&$T_{\rm{int}}$	&$\rm{rms}$\\
	&(B1950)    &(B1950)	&	&(keV)		&($10^{37}$W)&(Hours) 	&(Jy)	\\
\hline
Coma	&12 57 18.29&28 12 28.5	&0.0232	&$9.1\pm0.7$	&7.01			&80	&0.021	\\
A1795	&13 46 34.43&26 50 37.5 &0.0616	&$7.8\pm1.0$	&9.93	   		&115	&0.020	\\
A399	&02 55 05.33&12 50 57.6	&0.0715	&$7.0\pm0.4$	&6.78			&96	&0.030\\
A401	&02 56 12.55&13 22 50.1	&0.0748	&$8.0\pm0.4$		&11.76		&96	&(As A399)	\\
A478	&04 10 40.89&10 20 26.0	&0.0882	&$8.4^{+0.8}_{-1.4}$	&13.31		&74	&0.018	\\
A2142	&15 56 16.45&27 22 08.0	&0.0899	&$9.7^{+1.5}_{-1.1}$	&20.52		&73	&0.023	\\
A2244	&17 00 52.86&34 07 54.5	&0.0980	&$7.1^{+5.0}_{-2.2}$	&7.39		&91	&0.018	\\

\hline

\end{tabular}
\label{tab:clusters} 
\end{table*}


\subsection{Calibration and Data Reduction}
\label{sec:data_red}

The primary calibrator for all VSA observations is Jupiter.  We based
our calibration scale on the effective temperature of the planet at
34\,GHz: $T_{34}=155 \pm 5$\,K (\cite{Mason1999}).  The flux scale is
transferred to our other calibration sources: Cas A and Tau A.  The
calibrators are observed on a daily basis, allowing flux and phase
calibration at regular intervals.  Cas A and Tau A are partially
resolved on the longest VSA observations: we overcome this problem by
applying models as discussed in \cite{paper5}.  Full details of the
VSA calibration will be presented in a forthcoming paper.  Note that
in \cite{clive} and \cite{rafa} we re-scale our calibration to agree
with the recent WMAP results.

The data reduction pipeline for galaxy clusters is identical to that
employed in the processing of our CMB data, and is presented in detail
in \cite{paper1}.  Each observation is analysed independently using
the {\tt reduce} software, developed by the VSA team.  The procedure
is now highly developed, allowing virtually automatic correcting,
flagging, filtering and re-weighting of the data.  However, each raw
data file must be checked by eye at least once to eliminate some `bad'
data (due to bad weather or telescope malfunction), and to ensure
optimum quality in the reduced data.  It is also necessary to identify
files requiring special filtering depending on where the Sun, Moon or
a bright planet was during the observation.  The resulting calibrated
visibilities from each observation are taken and stacked together.

The data were reduced independently by the groups at the Cavendish,
the IAC and JBO, and the results found to be fully consistent.
Approximately 28$\%$ of the data were discarded due to bad weather,
filtering and telescope down--time.

The form of data from the single baseline source--subtraction
interferometer is identical to that of the main array and is
processed in a similar way. The primary flux calibrator is NGC 7027.
The flux scale from this is applied to our other flux calibrators.  We
use interleaved calibrators in order to monitor the telescope
phase.


\subsection{Radio Sources}
\label{sec:sources}

Contamination by radio sources can be a large problem for CMB
observations.  The contribution goes as $\ell^2$ so tends to be more
problematic for the (often higher-resolution) SZ work than for
primordial CMB observations.  In order to map the SZ effect
accurately, it is necessary to account for the effect of radio sources
which may be part of, in front of, or behind the cluster.  The VSA
source--subtraction interferometer allows potentially problematic
sources to be observed simultaneously with main array observations of
the cluster fields.

As no high frequency ($\approx$34\,GHz) survey of the radio sky is
available, we scheduled source observations via a two--fold approach:
\begin{itemize}
\item{The NVSS and GB6 catalogues (\cite{NVSS}, \cite{GB6}) were
examined for sources within a radius of 2$^{\circ}$ from the cluster
centres.  Source fluxes at 1.4 and 4.9\,GHz were used to perform a
simple extrapolation to 30\,GHz, thus making some prediction of the
approximate level of contamination in the SZ observations.  All
sources with predicted flux greater than 50\,mJy were selected for
observation with the VSA source--subtractor}.
\item{In order to account for flat or rising spectrum sources not seen
at the lower frequencies, the RT was used to survey the central square
degree of each field at 15\,GHz with the rastering technique described
by \cite{liz}.  Peaks $\gtrsim$20\,mJy in the raster maps were
recorded and the corresponding position list was added to the source
subtractor observing queue. This ensured that we accounted not only
for all potentially bright sources in the field, but also for fainter
sources which may have been present in the critical central regions of
the SZ fields.}
\end{itemize}
A summary of the source lists for all clusters is presented in Table
\ref{tab:sources}, including fluxes measured by the source--subtractor.
The 15\,GHz fluxes are those from RT pointed observations.  Whereas
for our primordial anisotropy work source fluxes were subtracted
directly from the visibilities, we choose here to use our measured
fluxes as priors in the Bayesian fitting software.  Due to telescope
malfunction at various stages during our observing schedule, not all
sources were observed simultaneously with the corresponding cluster.
In order to account for possible variability in the source flux,
broader priors were used than would have been assumed otherwise.
Directly subtracting source fluxes with such uncertainties would lead
to biases when fitting to the SZ data.

\begin{table*}
\centering

\caption{Radio sources present in the cluster fields.  The asterisked
source was predicted to have flux less than 50\,mJy, but
\protect\cite{Mason_obs} suggest it may be variable.}

\begin{tabular}{llccccc}
    \hline

        &RA  	 &Dec  	&Predicted Flux &RT Survey&VSA Source--Subtractor\\
	& & &34\,GHz &15\,GHz &34\,GHz\\
	&(B1950)  &(B1950)  	&(\,mJy) &(\,mJy)&(\,mJy)\\
\hline
Coma	&12 48 36 &+28 39 47	&75	&&$46\pm11$\\
	&12 49 25 &+28 07 55	&71	&&$29\pm9$\\
 	&12 50 49 &+27 55 57 	&99	&&$82\pm8$\\
	&12 51 46 &+27 53 41	&311	&&$250\pm3$\\
	&12 54 04 &+27 17 17	&57	&&$56\pm5$\\
	&12 55 36 &+28 36 36	&96	&$49\pm3$&$26\pm9$\\
	&12 56 08 &+29 25 19	&53	&&$10\pm12$\\
	&12 57 11 &+28 13 40	&-	&$27\pm3$&$34\pm7$\\
	&12 58 04 &+28 46 18	&226	&$251\pm13$&$207\pm10$\\
	&12 58 56 &+28 37 45	&-	&$34\pm3$&$31\pm5$\\
	&12 58 59 &+28 58 59	&168	&&$10\pm7$\\
	&12 59 58 &+27 25 17	&58	&&$49\pm9$\\
	&13 03 59 &+27 18 37	&52	&&$45\pm9$\\
\hline
A1795 	&13 39 50 &+27 24 42	&521	&&$380\pm9$\\
	&13 45 45 &+25 16 01    &521	&&$12\pm7$\\
	&13 46 09 &+26 42 42	&89	&&$8\pm10$\\
	&13 46 34 &+26 50 25	&36	&$51\pm3$&$31\pm9$\\
	&13 49 03 &+27 19 48	&-	&$8\pm3$&$20\pm11$\\
	&13 49 41 &+25 24 17	&71	&&$7\pm6$\\
\hline
A399/A401&02 53 51 &+13 22 25	&325 	&$342\pm17$&$235\pm8$\\
	&02 55 24 &+13 40 10	&32*	&&$36\pm4$\\	
	&02 55 47 &+13 22 19	&37	&$52\pm3$&$29\pm4$\\
	&02 56 01 &+11 31 00	&84	&&$54\pm9$\\
	&02 56 52 &+13 42 59	&35	&$66\pm3$&$26\pm5$\\
	&02 57 25 &+11 25 45	&60	&&$55\pm4$\\	
	&02 58 34 &+13 03 53	&28 	&$17\pm3$&$13\pm6$\\
	&02 59 48 &+12 07 18	&305	&&$107\pm9$\\
	&03 00 23 &+12 57 22	&80	&&$97\pm7$\\
\hline
A478	&04 08 52 &+08 35 38	&190	&&$61\pm12$\\
	&04 10 55 &+11 04 43	&836	&&$395\pm9$\\
	&04 11 02 &+10 10 19	&-	&$14\pm3$&$7\pm4$\\
\hline
A2142	&15 48 08 &+27 27 02	&166	&&$58\pm7$\\
	&15 52 28 &+27 55 35	&61	&&$2\pm6$\\
	&15 58 04 &+27 11 13	&163	&&$5\pm6$\\
	&15 58 57 &+26 53 35	&-	&$56\pm3$&$17\pm6$\\
	&16 00 03 &+26 18 43 	&57	&&$38\pm6$\\
	&16 00 35 &+26 54 15	&498	&&$176\pm14$\\
	&16 04 54 &+27 25 22	&326	&&$186\pm17$\\
\hline
A2244	&16 53 50 &+32 48 55	&88	&&$48\pm6$\\
	&16 56 12 &+34 48 01	&512	&&$297\pm11$\\
	&17 06 12 &+33 50 37	&110	&&$95\pm8$\\
\hline
\end{tabular}
\label{tab:sources} 
\end{table*}

We can assess how much the SZ detections are affected by confusion
noise from sources not found in the above, as follows.  A corollary of
Scheuer's work (\cite{scheuer}) is that confusion is worst when there
is $\approx1$ source per synthesised beam.  Examination of Table
\ref{tab:sources} shows that in the RT surveying, at about 20\,mJy
there is less that one source per VSA average SZ synthesised beam.  A
rough extrapolation indicates that there is one source per beam at
34\,GHz at a level of 10\,mJy.  Since the detected SZ fluxes are
$\approx150$\,mJy, it is evident that the source strategy is adequate.


\section{RESULTS}

\subsection{Maps}

The flagged and stacked data are held as visibility files, containing
the real and imaginary part for each observed $uv$--position along
with an associated rms noise level. Standard AIPS tasks are used to
make maps, and to perform CLEANing using one CLEAN box encompassing
the area of the VSA primary beam. All analysis and parameter fitting
is performed in the visibility plane; the maps presented here along
with the resulting discussion are included purely to illustrate the
results of our SZ programme.

\begin{figure*}
\begin{minipage}[b]{1.0\linewidth}
\centering\epsfig{file=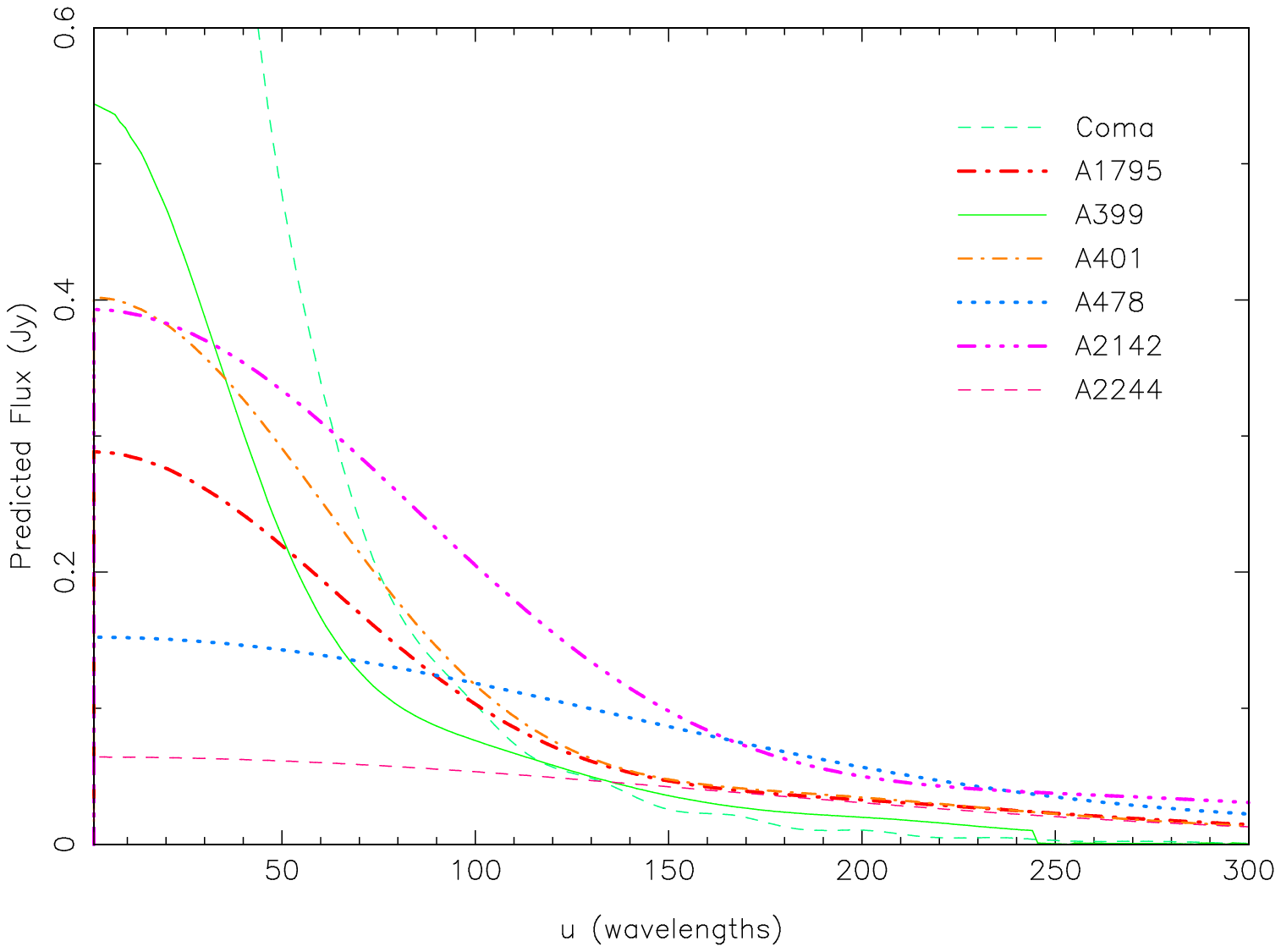,width=\linewidth,angle=0,clip=}
\end{minipage} 
\caption{Predicted SZ profiles for the cluster sample.}
\label{fig:profiles}
\end{figure*}

We expect a larger SZ response on the shortest baselines, so an
appropriate Gaussian taper is applied in each case.  This emphasises
structure on large scales.  Taper values were chosen based on the
range of $uv$ radii available in each cluster's data.  In order to
determine appropriate tapers for our sample, we used cluster
parameters from \cite{Mason_obs} (as listed in Table
\ref{tab:x-params}) to generate predicted SZ profiles.  These are
shown in Figure \ref{fig:profiles}.  (We observe that the
\cite{Mason_obs} value for the core radius of A399 ($4.33\pm0.45$ arcmin) 
is in direct conflict with that reported by \cite{bham} ($1.89\pm0.36$
arcmin).  The use of \citeauthor{Mason_obs}'s parameter may result in
an over-estimate of the SZ flux from this cluster.)  The chosen tapers
are $\approx0.1\rm{k}\lambda$, although the taper for Coma would
ideally be $\approx0.023\rm{k}\lambda$.  This cuts out nearly all
Extended Array baselines, so a value of $\approx0.1\rm{}k\lambda$ was
used with good results.  These maps of the VSA cluster sample are
presented in Figure \ref{fig:maps}.  The contours are $1.5\sigma$,
where $\sigma$ is the rms noise level presented in Table
\ref{tab:clusters}. We comment on the significance of the detections in
each map, and also the strength of the observed primordial features.
We emphasise that this is not intended to be a quantitative analysis
of the signal to noise ratio achieved for each cluster.

\subsubsection{Coma: Map (a)}
Coma is at redshift $z=0.0232$, giving it an angular size on the sky
roughly four times greater than any other cluster in the sample.  It
would ideally be observed on baselines even shorter than those of
the VSA.  However, the SZ signal from this cluster is so strong, we
detect it at 7.5$\sigma$.  4.5$\sigma$ primordial features are visible
around the SZ decrement.

\subsubsection{A1795: Map (b)}

A1795 is also detected at the 7.5$\sigma$ level.  This map 
contains a bright positive primordial feature south of the cluster.

\subsubsection{A399 and A401: Map (c)}

A399 does not appear in the map.  We argue that this is most probably
due to contamination by primordial CMB.  Although the contours are
negative at the position of A401, we suggest that this is largely due
to the primordial decrement east of the cluster position.  The SZ
signal from the cluster may be contributing in part, but it is
important not to confuse the two effects.  The centre of the obvious
decrement is around 15 arcmin away from the X-ray centre of A401.

\subsubsection{A478: Map (d)}

The A478 map shows a 6$\sigma$ SZ detection.  Primordial CMB
structures are visible all around the cluster, varying in strength
from 3--4.5$\sigma$.

\subsubsection{A2142: Map (e)}

The 7.5$\sigma$ detection of A2142 appears to be relatively free from
bright primordial features.

\subsubsection{A2244: Map (f)}

A2244 does not appear in the map.  Again, we suggest that
the cluster may be coincident with a peak in the CMB.
\\

\begin{figure*}
\begin{minipage}[b]{0.36\linewidth}
\begin{center}
\subfigure[]{
\includegraphics*[width=\linewidth]{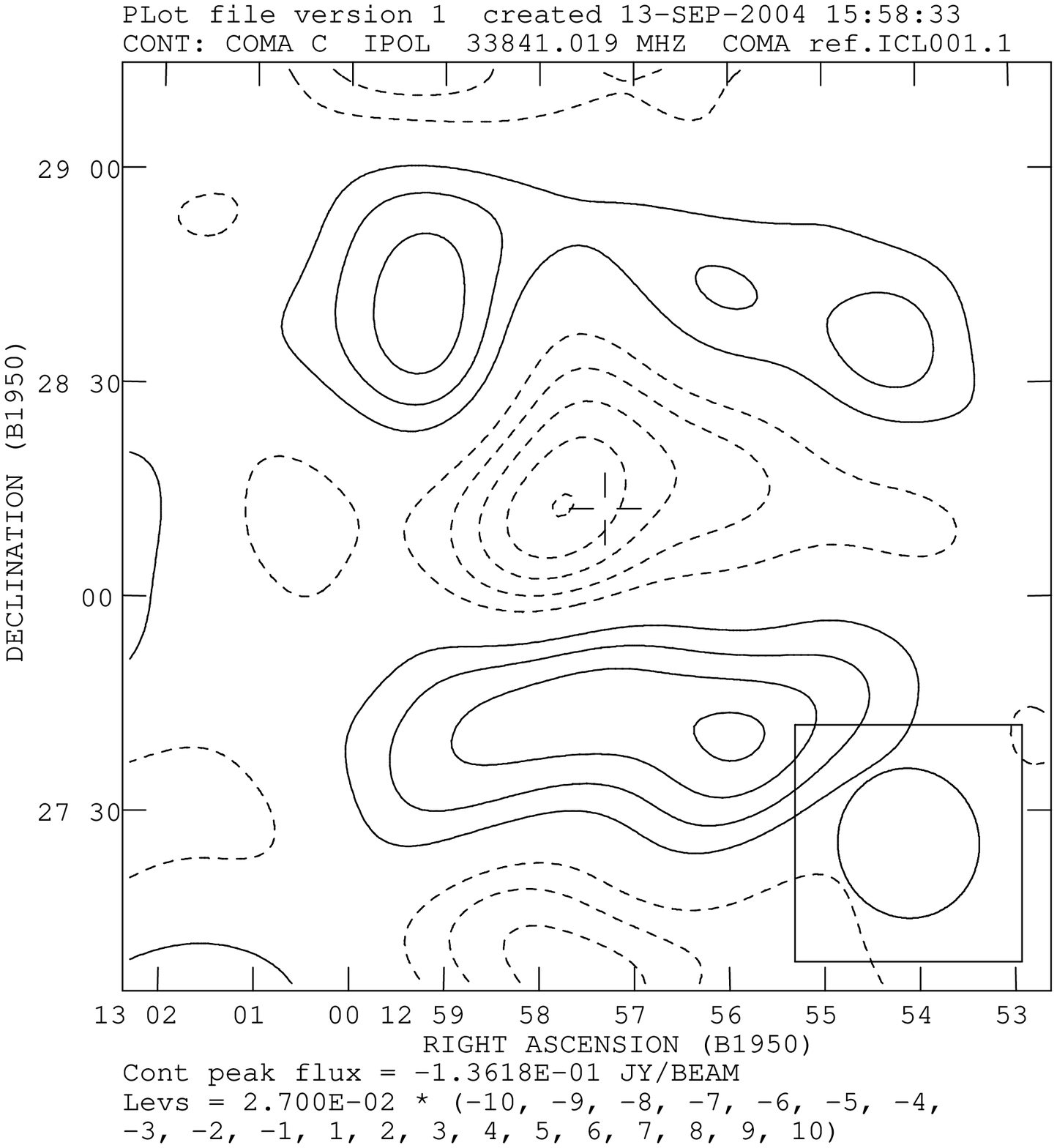}}
\end{center}
\end{minipage} 
\begin{minipage}[b]{0.36\linewidth}
\begin{center}
\subfigure[]{
\includegraphics*[width=\linewidth]{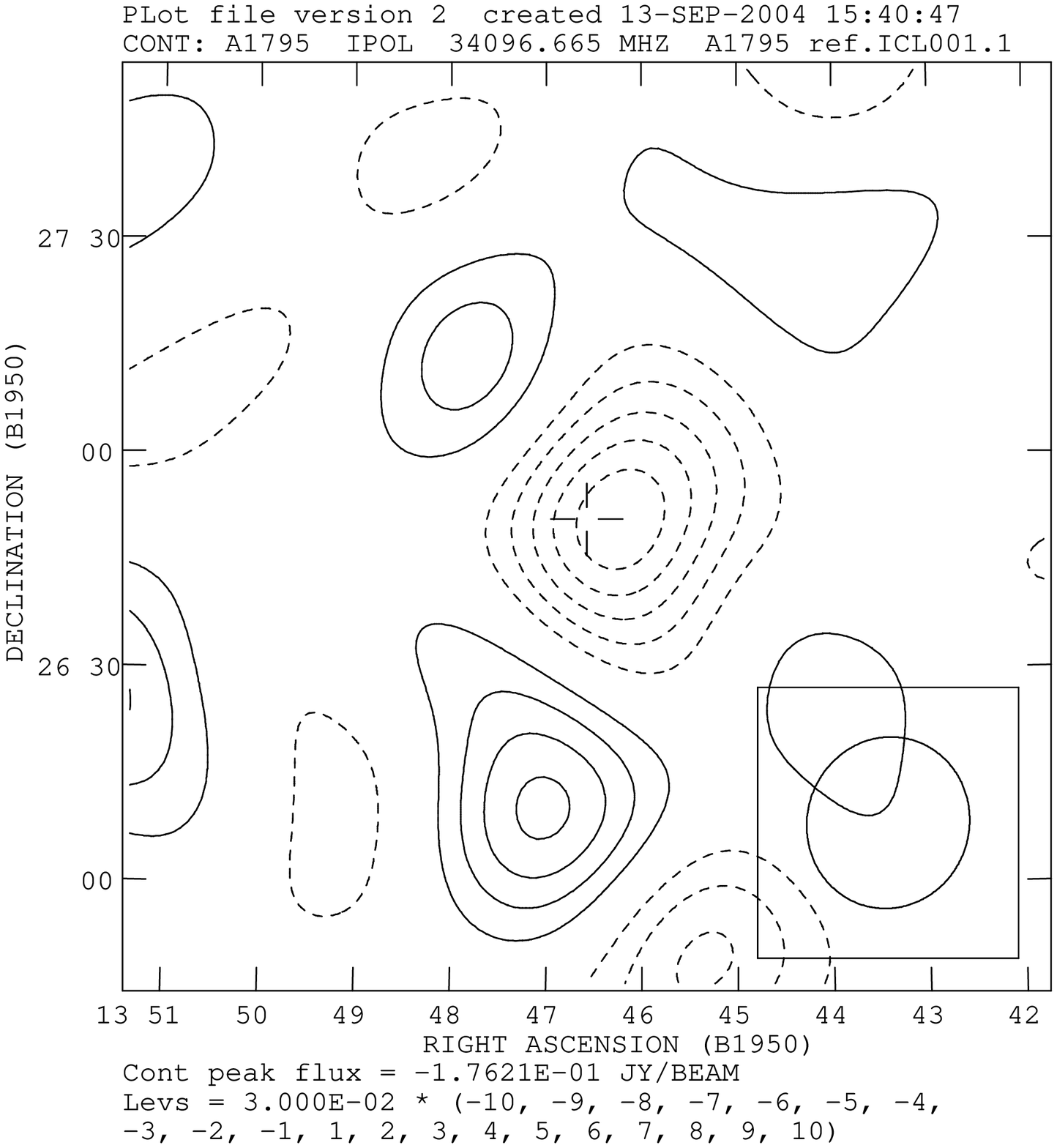}}
\end{center}
\end{minipage} 
\begin{minipage}[b]{0.36\linewidth}
\begin{center}
\subfigure[]{
\includegraphics*[width=\linewidth]{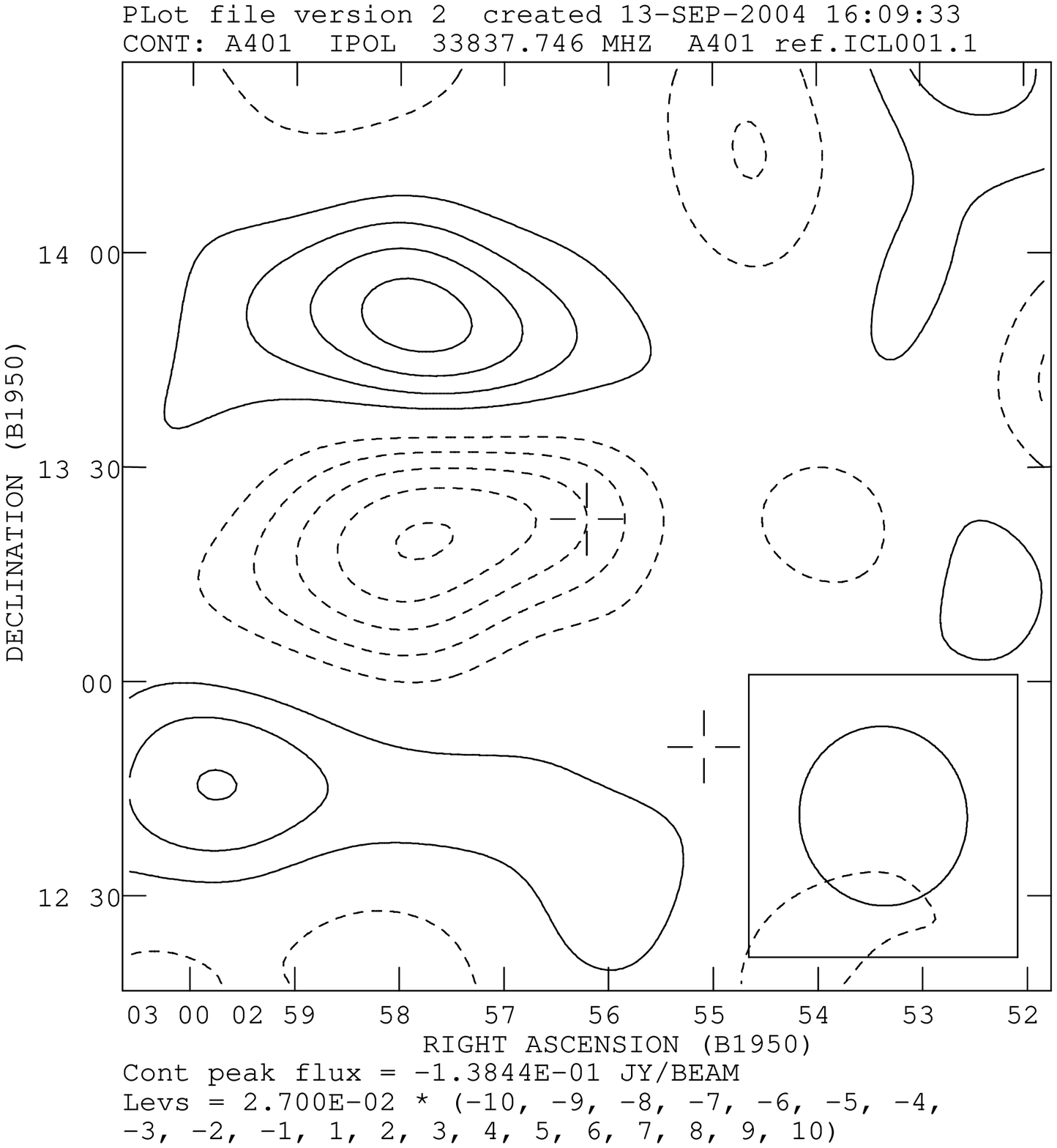}}
\end{center}
\end{minipage} 
\begin{minipage}[b]{0.36\linewidth}
\begin{center}
\subfigure[]{
\includegraphics*[width=\linewidth]{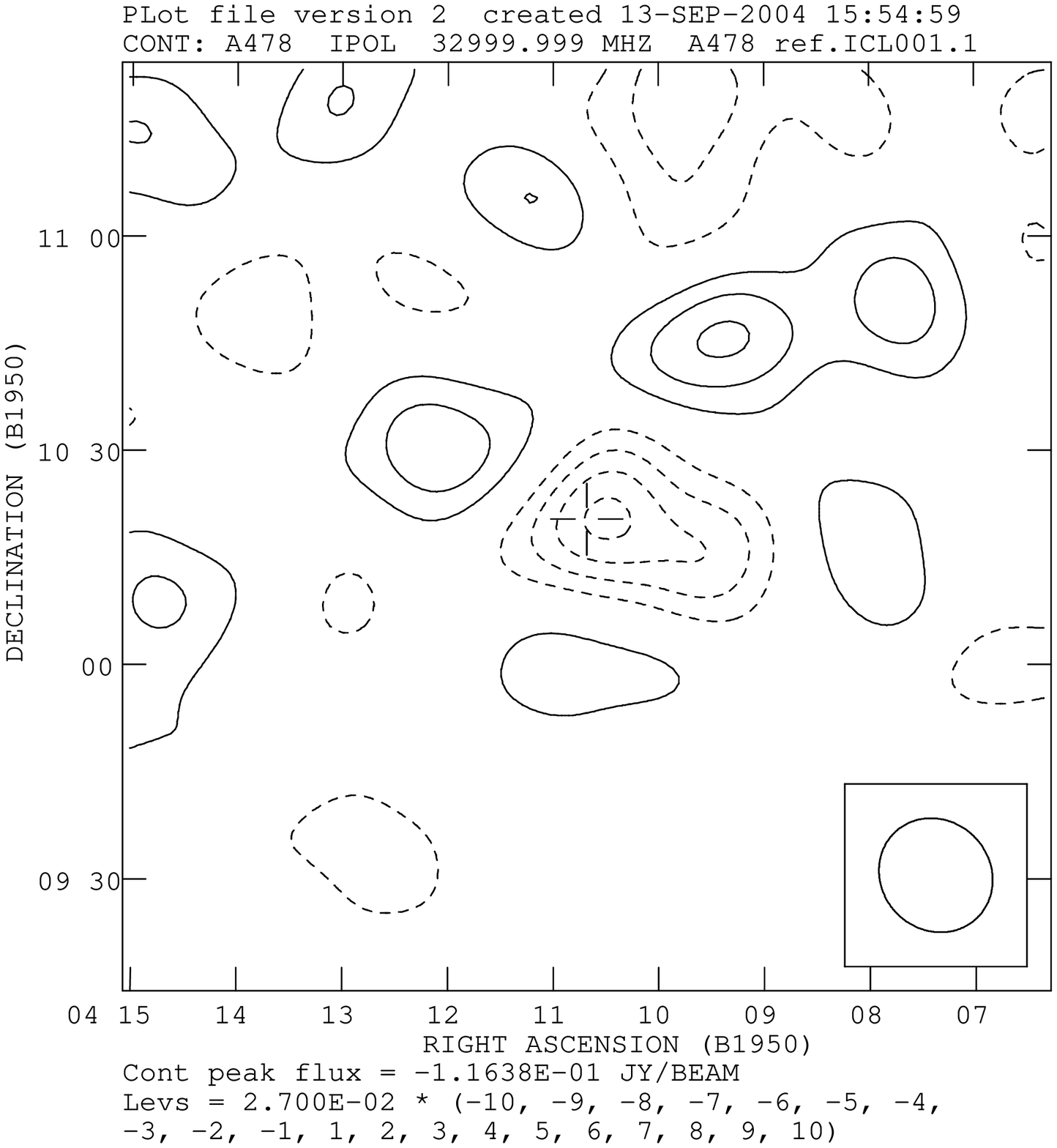}}
\end{center}
\end{minipage} 
\begin{minipage}[b]{0.36\linewidth}
\begin{center}
\subfigure[]{
\includegraphics*[width=\linewidth]{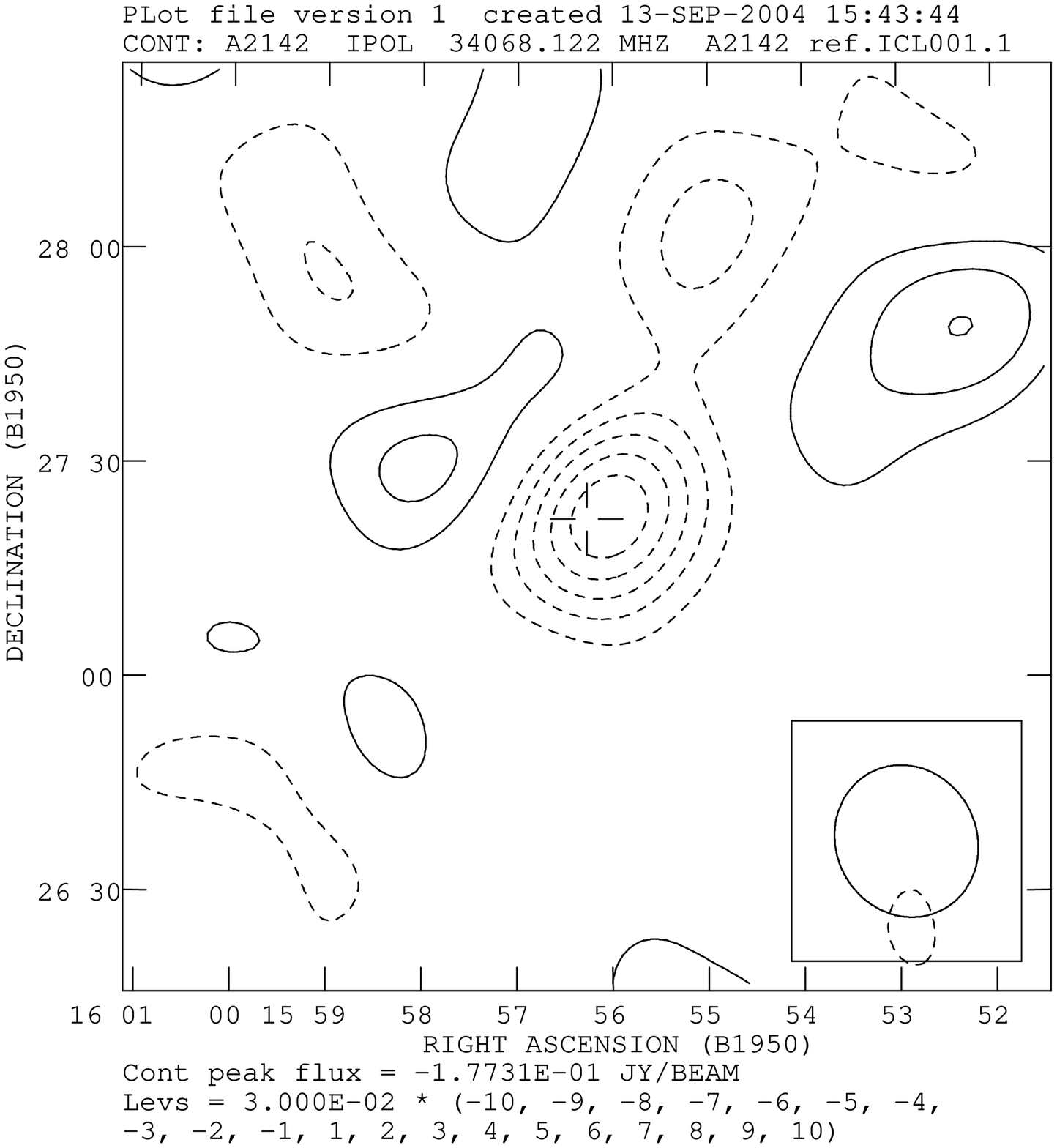}}
\end{center}
\end{minipage} 
\begin{minipage}[b]{0.36\linewidth}
\begin{center}
\subfigure[]{
\includegraphics*[width=\linewidth]{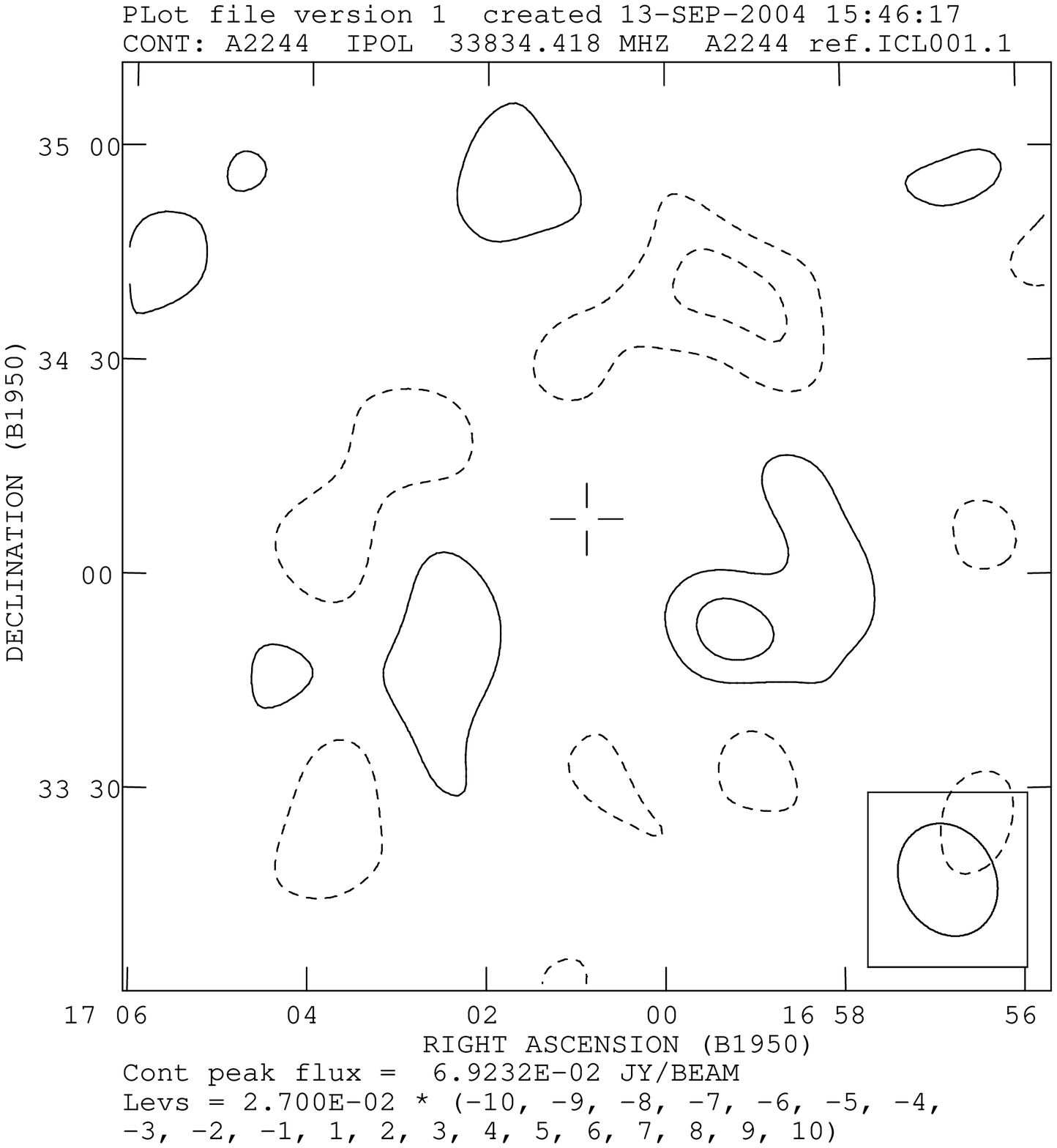}}
\end{center}
\end{minipage} 
\caption{CLEANed VSA maps ((a)--(f)) of the clusters Coma, A1795,
A399/A401 (where A399 is furthest south), A478, A2142 and A2244.  The
X--ray centre is marked in each case.  The half--power CLEAN beam is
shown in the bottom right corner of each plot, contours are
$1.5\sigma$. Radio sources have been subtracted and the coordinates
are B1950.}
\label{fig:maps}
\end{figure*}


\subsection{Cluster Model}
\label{sec:model}

In the SZ effect, incident CMB photons are Compton scattered by
the hot gas in a cluster's potential well.  At frequencies less than
217\,GHz, a brightness temperature decrement in the microwave
background is observed.  This is proportional to the `Comptonisation
parameter'

\begin{equation}
y = \frac{\sigma_{\rm{T}}}{m_{\rm{e}}c^2}\int{n_ekTdl},
\label{eqn:y}
\end{equation}

\noindent which is proportional to the line integral of pressure
through the cluster.  This can be calculated from modelled gas density
distributions.

As we are working with specifically large--angular scale SZ data,
contamination from primordial CMB features is considerable, thus
adding an extra `noise' term.  (In our parameter inference, this is
dealt with appropriately as an additional source of Gaussian noise -
see section \ref{sec:contaminants}). This restricts us to a highly
constrained, simple model.  We choose to follow \cite{grego} in
fitting a $\beta$--model (\cite{beta1},
\shortcite{Cavaliere1978}) to the cluster visibilities. We too simplify the
problem by assuming the clusters to be spherically symmetric and in
hydrostatic equilibrium. (Note: Strictly the assumptions of
isothermality, $\beta$--profile, and hydrostatic equilibrium are
incompatible.  However, to good approximation, they are compatible
over a wide range of $r$ for $\beta$ close to 2/3. See
\cite{King1962}.) In the $\beta$--model, the gas density as a function
of radius takes the form

\begin{equation}
\rho_{\rm gas}(r) = \frac{\rho_{\rm
gas}(0)}{\left(1+\left(r/r_{\rm{c}}\right)^{2}\right)^{\frac{3\beta}{2}}},
\label{eqn:betaprofile}
\end{equation}  

\noindent where $r_{\rm{c}}$ (core radius) and $\beta$ are parameters
of the fit.  From the assumptions of hydrostatic equilibrium and gas
isothermality at temperature $T$,

\begin{equation}
\frac{kT}{\mu}\frac{d\rho_{\rm{gas}}}{dr} =
-\rho_{\rm{gas}}\frac{GM_{\rm{r}}}{r^{2}},
\label{eqn:hyd}
\end{equation}

\noindent where $M_{\rm{r}}$ is the total mass internal to $r$, $\mu$ is the
mass per particle, and $k$ and $G$ are the Boltzmann and gravitational
constants.  Equations (\ref{eqn:betaprofile}) and (\ref{eqn:hyd}) lead
to the following expression for the total mass distribution:

\begin{equation}
M_{\rm{r}} = \frac{3\beta r{^3}}{(r_{\rm{c}}^{2}+r^{2})}\frac{kT}{\mu G}.
\label{eqn:4}
\end{equation}

\noindent This can be adapted usefully to
calculate cluster masses out to some overdensity, e.g. $r_{200}$.

\begin{eqnarray}
M_{200} &=& \frac{4\pi}{3}r_{200}^{3}(200\rho_{\rm{crit}}) \\ 
&=& \frac{3\beta
r_{200}^{3}}{(r_{\rm{c}}^{2}+r_{200}^{2})}\frac{kT}{\mu G}
\label{eqn:MB}
\end{eqnarray}

\noindent In this work, we choose to calculate quantities out to
$r_{200}$ as this is a good approximation to the virial radius of a
cluster.  Previous studies have used $r_{500}$ so we have also
extended our calculations to produce results to this radius for
comparison purposes.

From the gas density distribution (\ref{eqn:betaprofile}) it is
straightforward to compute the gas mass to this radius:

\begin{eqnarray}
M_{\rm{gas}} &=& \int_{0}^{r_{200}}{4\pi r^{2}\rho_{\rm{gas}}dr}\\
&=&4\pi \rho_{\rm{gas}} (0) r_{\rm{c}}^{3} \int_{0}^{\frac{r_{200}}{r_{\rm{c}}}}
\frac{x^{2}dx}{(1+x^2)^{\frac{3\beta}{2}}}.
\label{eqn:8} 
\end{eqnarray}

\noindent The above integral is evaluated numerically.  We choose to
parameterise in terms of $M_{\rm{gas}}$, and can solve for the
gas density in order to compute the Comptonisation parameter.  The
calculated values can then be compared to real VSA data.

The gas fraction is defined as

\begin{equation}
f_{\rm{gas}} = \frac{M_{\rm{gas}}}{M_{\rm{r}}},
\label{eqn:f_gas}
\end{equation}

\noindent in which $M_{\rm{gas}}$ and $M_{\rm{r}}$ are evaluated to
the same radius.  $f_{\rm{gas}}$ evaluated by this method is
proportional to $h^{-1}$.  One way to see this is as follows.  In
equation \ref{eqn:8}, the $h$--dependences of the limit
$r_{200}/r_{\rm{c}}$ cancel, $\rho_{\rm{gas}}(0)$ is a local quantity
and so not $h$--dependent, and only $r_{\rm{c}}^2$ depends on $h$
because the third factor of $r_{\rm{c}}$ is along the line of sight;
thus $M_{\rm{gas}} \propto h^{-2}$.  In equation \ref{eqn:4},
$M_{\rm{r}} \propto r^3/(r_{\rm{c}}^2+r^2) \propto h^{-1}$. So,
$f_{\rm{gas}} \propto h^{-1}$.


\subsection{Interferometric Data}
\label{sec:int_dat}

Interferometers sample the $uv$--plane so it follows that the most
straightforward approach is to fit to the visibility data directly.
This is further motivated by the following points.  The instrument
noise is Gaussian in the $uv$--plane, and independent between
visibilities. In the map plane the noise is highly correlated
spatially.  In addition, fitting to the visibilities naturally avoids
the problem of synthesised beam deconvolution.  The primordial CMB is
well understood in the $uv$--plane in terms of the measured power
spectrum, so can be factored into the computation (see
\ref{sec:contaminants} for details).  Finally, the inclusion of point
sources is straightforward.


\subsection{Contaminants}
\label{sec:contaminants}

There are two relevant astrophysical contaminants to the SZ data:
primordial fluctuations in the CMB, and foreground radio sources.
Emission from the Galaxy is taken to be negligible in this analysis.

Primordial CMB fluctuations, recognised as a source of Gaussian noise
with known angular power spectrum, are included in a non-diagonal
covariance matrix when calculating the misfit between predicted and
observed data (\cite{Reese}, \cite{phil1}).  We observed bright
primordial features in all of our cluster maps, and indeed they are
evident in Figure \ref{fig:maps}.  As the negative primordial features
are of similar strengths and on similar angular scales to the cluster
decrements, it is necessary to apply fairly tight positional priors
(see section \ref{sec:par_inf}).  As regards $f_{\rm{gas}}$ estimates,
we argue that the position is acceptable as the effect of the CMB
tends to produce a cancelling effect on $M_{\rm{gas}}$ and
$M_{\rm{r}}$ (see section \ref{sec:primordials}).

The point sources present in each field are also included in the model
of the sky. The source--subtractor data allow the determination of the
fluxes and positions of these objects: we translate these measurements
into appropriate priors (see section \ref{sec:par_inf}) on the source
parameters.  These `nuisance parameters' are then marginalised out.


\subsection{ Parameter Inference}
\label{sec:par_inf}

\subsubsection{Basic considerations}

In inferring cluster parameters, the traditional route followed in the
literature is the Maximum Likelihood method. This method was used
in, for example, the SZ and gas fraction work of \cite{grego}.
Computational restrictions at the time prevented the use of the fully
Bayesian analysis we perform in this paper.  The likelihood of a
dataset $\rm{L}(data|\boldsymbol{\theta})$ is the product of the
probability distributions of the constituent data points, where
$\boldsymbol{\theta}$ is used to characterise a set of parameters such
as $\beta$ and core radius.  This likelihood may be maximised to find
the best--fit value for each parameter of the set
$\boldsymbol{\theta}$.  This approach:
\begin{enumerate}
\item{assumes that the parameters $\boldsymbol{\theta}$ of a model
have a true set of values, and that obtaining data from an appropriate
experiment will measure this set of values;}
\item{can be formulated in terms of a single misfit statistic when
describing the difference between the predictions of a model and a
measurement: maximising a Gaussian likelihood for data with
uncorrelated errors is equivalent to minimising the mean--squared
residual, or chi-squared statistic;}
\item{usually assumes Gaussian noise, although indeed this can be
modified to incorporate the correct distribution (e.g. Poisson) for a
particular case.}  
\end{enumerate}

The Maximum Likelihood method focuses on the estimation of true
parameters from data, while neglecting the full distributions for
those parameters.  When signal--to--noise is low, these distributions
are broad and very unlikely to be Gaussian: we summarise the
difficulties in this situation as follows.

Maximum Likelihood does not describe the joint process of observation
and inference.  We have a set of noisy visibilities (the data) which
we attempt to explain by a model or hypothesis, H.  The hypothesis
includes the notions, for example, that the SZ signal comes from a gas
distribution (which we assume here to have a $\beta$--profile) and that
sources and CMB primordials are present, and also the assumption that
we understand the experiment in question (i.e. the interferometer
works).  The data model includes the parameter set
$\boldsymbol{\theta}$ as defined above.  We wish to estimate
$\boldsymbol{\theta}$ from our data, that is, we wish to examine the
probability distribution
$\rm{P}(\boldsymbol{\theta}|\rm{data},\rm{H})$.  N.B.: the notation
$\rm{P}(A|B)$ refers to the probability of A given B.  Rather than
achieving this, the Maximum Likelihood method assesses the data while
taking it as \emph{given} that $\boldsymbol{\theta}$ has some true
value, as outlined in point (i) above.  In other words, it evaluates
just the peak of the probability distribution
$\rm{P}(\rm{data}|\boldsymbol{\theta},\rm{H})$.  Application of Bayes'
theorem allows us to relate the two distributions
$\rm{P}(\boldsymbol{\theta}|\rm{data},\rm{H})$ (the \emph{posterior})
and $\rm{P}(\rm{data}|\boldsymbol{\theta},\rm{H})$ by

\begin{equation}
\rm{P}(\boldsymbol{\theta}|\rm{data},\rm{H}) =
\frac {\rm{P}(\rm{data}|\boldsymbol{\theta},\rm{H})
\rm{P}(\boldsymbol{\theta}|\rm{H})} {\rm{P}(\rm{data}|\rm{H})}.
\label{eqn:bayes}
\end{equation}
  
The additional factors in equation \ref{eqn:bayes} are the
\emph{prior} probability distribution,
$\rm{P}(\boldsymbol{\theta}|\rm{H})$, and the \emph{evidence},
$\rm{P}(\rm{data}|\rm{H})$, to which we will return shortly.

In addition, point (ii) is not generally correct. Even if
$\rm{P}(\rm{data}|\boldsymbol{\theta}, H)$ is Gaussian, it is
multiplied by the prior $\rm{P}(\boldsymbol{\theta}|\rm{H})$ which
may, for example, be asymmetric.  Once one starts to produce resultant
probability density functions by multiplication the distributions are
certainly going to be complicated.  The probabilities outlined above
are \emph{functions}. The standard Maximum Likelihood approach
characterises such probability distributions by a single value with an
error bar.  The characterisation of probability distributions with
approximate Gaussians is therefore misleading and may underestimate the
final uncertainty in a quantity such as $\rm{f}_{\rm{gas}}$. It is
clearly preferable to retain all the information contained in the
entire function, rather than working with
single--value parameters.  As mentioned above, point (iii) can be
dealt with appropriately.

Propagating the likelihood function via Bayes' theorem thus overcomes
points (i) and (ii) above.  It also delivers additional advantages,
summarised as follows:

\begin{itemize}
\item{Conditioning on a particular value of a parameter implies a
delta--function prior, a state of knowledge that never occurs.  It is
now possible to deal with continuous probability distribution
functions in many dimensions (e.g. positions, core radii, $M_{\rm{r}}$
etc.)  rather than having to work just with peaks and widths of
artificially low--dimension probability distributions.  A desire to
concentrate on a subset of interesting parameters leads directly to
the concept of marginalisation (see e.g. \cite{Sivia}) }.
\item{The method leads directly to the evaluation of the evidence, an
extremely useful quantity that enables one to assess the relative
suitability of a set of hypotheses (see e.g. \cite{COS/HBL02}).}
\end{itemize}

The evidence in Equation (\ref{eqn:bayes}) is $\rm{P(data|H)}$ and is
an integral over all parameters in N--dimensional parameter vector
$\boldsymbol{\theta}$:

\begin{equation}
\rm{P(data|H)} = \int
\rm{P(data|}\boldsymbol{\theta},\rm{H})\rm{P}(\boldsymbol{\theta}|\rm{H})
\rm{d}^{\rm{N}}\boldsymbol{\theta}
\label{eqn:evidence} 
\end{equation}

This can be applied usefully to help distinguish between different
hypotheses, say $\rm{H}_{1}$ and $\rm{H}_{2}$:  Bayes' theorem
(equation \ref{eqn:bayes}) can be applied in order to evaluate and
compare $\rm{P(H_{1}|data)}$ and $\rm{P(H_{2}|data)}$.  In doing this,
$\rm{P(data)}$ cancels out and we obtain

\begin{equation}
\frac{\rm{P(H_{1}|data)}}{\rm{P(H_{2}|data)}} =
\frac{\rm{P(data|H_{1})}}{\rm{P(data|H_{2})}} \frac{\rm{P(H_{1})}}{\rm{P(H_{2}) }}
\label{eqn:hyp}
\end{equation}

Thus hypotheses may be compared.  For example, we can evaluate the
hypothesis that an SZ cluster is in a particular, small patch of sky.
We can compare this with the evidence given an alternative hypothesis,
this time deeming that the cluster be found in a larger area of sky.
The hypothesis probability ratio given in equation \ref{eqn:hyp}
provides the means by which the suitability of these two priors can be
assessed.  Such additional information may be obtained from elsewhere;
in this particular example X-ray data may be used to good effect.

We note that both Maximum Likelihood and Bayesian methods can cope
with correlated data (See e.g. \cite{phil1}, \cite{Reese} as before)
but simple chi-squared minimisation cannot.

\subsubsection{Characterising the posterior Probability Density
Function (PDF)}
\label{sec:par_inf:mcmc} 

Having summarised the advantages of the Bayesian route, we now turn to
the problem of calculating the posterior distribution
$\rm{P}(\boldsymbol{\theta}|\rm{data, H})$.  One method is to evaluate
it as a product of the probabilities for every visibility, for all
possible values of each of the N parameters in $\boldsymbol{\theta}$.
This is the `brute force' approach, involving the calculation of the
likelihood over a huge hypercube.  This technique is now plausible for
application to the CMB primordial power spectrum, given that the CMB
itself has a Gaussian brightness probability distribution at every
point on the sky (and is indeed the same everywhere).  However, it is
not a realistic approach for an SZ $\beta$--model with position, mass
and size uncertainties in the presence of the CMB and a number of
radio sources.  So we have chosen to represent the posterior in an
approximate way by drawing samples from it, the Markov Chain Monte
Carlo method (see e.g.  \cite{MCMC}, \cite{Ruanaidh} for general
introductions, and \cite{phil1}, \cite{Bonamente2004} for galaxy
cluster specifics).

This process results in a set of sample parameter vectors whose number
density is proportional to the posterior probability, such that all
local maxima are explored in proportion to their relevance. In order
to ensure that the correct regions of parameter space are being
probed, sufficient samples must be taken and calculations made.  This
is problematic in that it must be both accurate and efficient: to this
end, we use the commercially available sampler `\bayesys'
(\cite{bayesys3}), a powerful code designed to be flexible enough to
cope with a wide range of problems.  \texttt{BayeSys} makes use of a
range of proposal distribution `engines' that govern where next to
sample, and in particular employs those that it finds dynamically to
be most efficient for a particular posterior pdf. In addition, it should be possible to assess whether
or not enough evaluations have been performed over an acceptable range
of $\boldsymbol{\theta}$, that is when the process has `burnt in'.  A
review of such tests is given in \cite{cowlescarlin}.  We follow
\cite{phil1} and argue that several short, independent burn--ins are a
good idea to check that they agree.  The diagnostic we use is the
evidence itself, which we calculate by `Thermodynamic Integration'
(see e.g. \cite{Ruanaidh}). The method works as follows.  The evidence
(as given in equation \ref{eqn:evidence}) is

\begin{equation}
\rm{P(data|H)} = \int
\rm{P(data|}\boldsymbol{\theta},\rm{H})\rm{P}(\boldsymbol{\theta}|\rm{H})
\rm{d}^{\rm{N}}\boldsymbol{\theta} \equiv \rm{E(1)}
\label{eqn:e1}
\end{equation}.

\noindent We now write down

\begin{equation}
\rm{E}(\lambda) = \int
\rm{P}^{\lambda}(\rm{data}|\boldsymbol{\theta},\rm{H})
\rm{P}(\boldsymbol{\theta}|\rm{H})
\rm{d}^{\rm{N}}\boldsymbol{\theta}
\end{equation}.

\texttt{BayeSys} allows the running in parallel of several Markov
chains (typically 10 in our case).  The key to the method is as
follows.  The sampling starts with $\lambda=0$.  This means that the
new data are initially ignored with samples just being drawn from the
prior.  At this stage, remote regions of parameter space (that are at
least allowed by the prior) are sampled.  $\lambda$ is then gradually
raised to one, at a rate balancing the needs for computational speed
and accuracy in the log evidence calculation. The latter can be shown
to reduce to the numerical integral of the ensemble--averaged
log--likelihood with respect to $\lambda$ (\cite{Ruanaidh}).

\subsubsection{Practicalities}

It is always of utmost importance to ensure that one does not
over--interpret the data available.  This is crucial here, as we have
not only fairly noisy data (due to the faint nature of the effect
being studied), but also considerable contamination from point sources
and primordial CMB fluctuations.  As is evident in the VSA data
(figure \ref{fig:maps}), and previously mentioned in section
\ref{sec:contaminants}, CMB features may be comparable in strength to
the SZ decrement itself.  It would be quite possible to fit,
accidentally, to a negative CMB feature which would be very
misleading.  Our method avoids this danger by including all
contaminants in the model, and fitting all parameters simultaneously.
We have chosen to fit a simple but well--motivated model to our data,
but even so we must fit six parameters plus source fluxes and
positions.  This makes the task computationally expensive (vastly more
so than using Maximum Likelihood).  In order to extract parameters for
a single cluster, around 100 hours of computer time is required
(2\,GHz processor).  We do not expect to place tight constraints on,
for example, $\beta$ or $r_{\rm{c}}$ and we anticipate broad
probability distributions for all parameters.  However, when we
marginalise properly over all parameters we find some interesting
precisions on $f_{\rm{gas}}$.

In order to compare a sample model with the VSA data, we project the
model gas pressure and map the Comptonisation onto a grid.  A Fast
Fourier Transform is then performed, and interpolated onto the
$u-v$ coordinates.  These predicted visibilities are then compared
to the observed cluster visibilities.  Working directly with the
visibilities has the advantages described in section
\ref{sec:int_dat}.  We deal with point sources and the CMB in the
following natural way.  The Fourier transform of a delta function is a
constant amplitude sine wave.  This can be used to increment all the
predicted visibilities by a factor specific to each source's sample
parameters. The uncertainty on each measured visibility is Gaussian
and has contributions from both the thermal noise in the receivers
(which is uncorrelated) and the primordial CMB fluctuations (which are
correlated between adjacent points in the $u-v$ plane). The resultant
noise covariance matrix~$\mathsf{C}$ is non-diagonal but calculable
given a primordial power spectrum, assumed to be well known. The
likelihood of the visibility data is therefore
\begin{equation}
\rm{P(\boldsymbol{d}|}\boldsymbol{\theta},\rm{H}) =
\frac{1}{(2\pi)^{N_{\rm vis}}|\mathsf{C}|^{1/2}}
\exp{\left[-(\boldsymbol{d}-\boldsymbol{d_p})^{T}\mathsf{C}^{-1}(\boldsymbol{d}-\boldsymbol{d_p})\right]},
\end{equation}
where~$\boldsymbol{d}$ and~$\boldsymbol{d_p}$ represent the observed and
predicted visibility vectors respectively, and $N_{\rm vis}$ is the
number of visibilities.

\begin{table}
\centering

\caption{Priors for the cluster analysis.  Positions and gas
temperatures for individual clusters are quoted in Table \ref{tab:clusters}.} 

\begin{tabular}{ll}
    \hline
Parameter	&Prior \\
\hline
Position	&Gaussian, 1 arcmin\\
$r_{\rm{c}}$	&Uniform, 1--1000kpc \\
$\beta$		&Uniform, 0.3--1.5 \\
$T_{\rm{e}}$	&Gaussian, ASCA value $\pm15\%$\\
$M_{\rm{gas}}$	&Uniform, $(0.01-3.00)\times10^{14}$\\
\hline
\end{tabular}
\label{tab:priors} 
\end{table}

The priors used to characterise the various model parameters are
summarised in Table \ref{tab:priors}. As mentioned in section
\ref{sec:contaminants}, tight priors were placed on both the cluster
position, and point source positions and fluxes.  For the cluster
centroid, the X-ray centre (\cite{NORAS}) was included as a Gaussian
prior of width 1 arcmin. We chose to place a weak prior on core radius
such that it be determined by the data to hand.  The prior on the $\beta$
parameter encompasses the extremes of the range of values found in
clusters to date.  The temperature prior allows a generous error on
the fit.  Note that $f_{\rm{gas}}$ depends on $T^2$ -- see
\cite{grego}. The prior on the gas mass more than encompasses the
accessible range.  The point source fluxes included in the model were
also assigned Gaussian priors, based on the source--subtractor
measurements and their uncertainty.  The prior on each source flux was
broadened to account for variability of a factor of 1.33 times the
measured flux: this step was only taken when the epoch of the source
measurement was significantly different from that of the cluster
observation.  For the sources selected using predictions from lower
frequencies, positional accuracies were taken from the GB6
catalogue.  The sources detected in the RT surveys were
assumed to have positional uncertainty of $\pm40$ arcsec in both RA
and Dec; this is wide enough to cover even the weakest sources.


\subsection{The Effect of Primordials on $f_{\rm{gas}}$ Estimates}
\label{sec:primordials}

In the context of large angular scale SZ observations, the CMB is
additional noise which will provide a source of error in the
determination of $f_{\rm{gas}}$. This extra noise was dealt with
correctly when calculating cluster parameters (see Section
\ref{sec:contaminants}). However, here we present a simple argument
describing why, in situations where the SZ data is used to infer
\emph{both} the gas mass \emph{and} the total mass (as discussed in
\ref{sec:model}), the contamination is not as catastrophic as one may
anticipate. With the present data quality, fitting a $\beta$--model is
doing little more than fitting an offset plus a slope.  If there is
more negative signal due to a negative CMB feature coinciding with the
cluster position, then the $M_{\mathrm{gas}}$ estimate will be higher.
(NB: This is a simplistic argument because of course the contribution
to the Comptonisation parameter depends on the mass distribution which
is linked to the total mass.)  Now, in estimating $M_{\mathrm{r}}$,
the effect of the above will be to increase the central concentration,
increasing $\beta$ or decreasing $r_{\mathrm{c}}$. Examination of
Equation \ref{eqn:8} shows that this effect will increase an
estimation of $M_{\mathrm{r}}$.  So, in this type of scenario, as both
$M_{\mathrm{gas}}$ and $M_{\mathrm{r}}$ will be higher, the effects of
the CMB tend to cancel out when calculating $f_{\mathrm{gas}}$ for the
cluster in question.  A similar effect is observed for a bright
primordial feature - the SZ signal will tend to decrease, and $\beta$
will also decrease as the cluster will appear to be less centrally
condensed.  Thus, if the primordial CMB contamination happens to be
correlated over the measured $u$--range, then the effects on
$M_{\mathrm{gas}}$ and $M_{\mathrm{r}}$ tend to cancel, leaving
$f_{\mathrm{gas}}$ little affected.

In general, depending on the actual sizes, shapes and positions of the
primordial features behind the SZ decrement, $f_{\mathrm{gas}}$ may be
pushed higher or lower, or remain relatively unaffected as outlined
above.  Of course, if there is a Universal value of
$f_{\mathrm{gas}}$, then combining the results from a reasonable
number of clusters will both help reduce any remaining effects and
also help to evaluate the effect's magnitude.  One may
intuitively regard the cases presented above to be the `worst case
scenario', when in fact they appear not to cause too great a
difficulty.  

We have performed a simple simulation in order to examine this
cancelling effect semi--quantitatively.  Using our A478 data, we
placed a test source of flux $S_{\mathrm{add}}$ at the pointing centre
and re--calculated $f_{\mathrm{gas}}$.  Results for test sources in
the range $-100 > S_{\mathrm{add}} > 100$\,mJy are presented in Table
\ref{tab:sim}.  Although this is by no means a rigorous test of the argument
 postulated, we note that the values of $f_{\mathrm{gas}}$ for all
 $S_{\mathrm{add}}$ are consistent within errors.  This indicates that
 in this context (ie for our $uv$--range and chosen cluster sample),
 the effect of the CMB tends to cancel out in this context.  Note that
 typical SZ fluxes are $\approx$150\,mJy, whereas CMB plus receiver
 noise will typically produce features of $\approx$100\,mJy, and
 occasionally $>$150mJy.  From these simple calculations, we argue that
 estimations of $f_{\mathrm{gas}}$ should be relatively unaffected by
 the presence of primordial CMB in all but the worst cases.

\begin{centering}
\begin{table}
\caption{Gas fraction estimations for A478 with the inclusion of a
test contaminant source of flux $S_{add}$ at the cluster centre.  }
\setlength{\extrarowheight}{3pt}
\begin{tabular}{cr@{}l}
    \hline

$S_{\rm{add}}$	&\multicolumn{2}{c}{$f_{\rm{gas}}$}\\
(mJy)		&&		\\
\hline
-100		&0.056&${}^{+ 0.088}_{- 0.041}$\\
-50		&0.10&${}^{+ 0.12}_{- 0.06}$\\
-25		&0.12&${}^{+ 0.14}_{- 0.07}$\\
0		&0.12&${}^{+ 0.11}_{- 0.06}$\\
25		&0.13&${}^{+ 0.14}_{- 0.07}$\\
50		&0.11&${}^{+ 0.13}_{- 0.06}$\\
100		&0.10&${}^{+ 0.09}_{- 0.05}$\\
\hline
\label{tab:sim}
\end{tabular}
\setlength{\extrarowheight}{0pt}
\end{table}
\end{centering}


\subsection{Other Effects on $f_{\mathrm{gas}}$}

In this work, the random errors present are larger than any
systematics, but here we present a brief discussion of some possible
additional sources of error.  Our assumptions of isothermality and
sphericity may affect our inferred values for $f_{\mathrm{gas}}$.  If
a cluster were not isothermal, we may, for instance, overestimate the
temperature in the outer regions due to a temperature gradient, and
may overestimate both the gas and total mass with a possible small net
underestimate of the gas fraction.  Regarding asphericity, which we do
not expect to have a large effect since we are not using X-ray surface
brightness, we point out that our sample is orientation unbiased,
because our flux limit is well above the flux limit of the X-ray
survey from which the clusters were chosen. \cite{grego} made mock
observations of a simulated cluster population, finding no bias as a
result of using a spherical isothermal $\beta$--model, suggesting that
these two sources of systematic error indeed may not be significant in
this work.  Additionally, \cite{Arnaud2004} find that the temperature
variation for clusters observed with XMM-Newton is less than 10$\%$
out to half the virial radius, and similarly \cite{Zhang2004} find
errors on mass estimates from XMM-Newton data to be less than $25\%$
as a result of temperature gradients.  Generally, X-ray derived
pressure maps seem to show a factor of two less variation, for example
azimuthally around the cluster centre, than either density or
temperature maps.  Still, gas clumping could be a problem. Clumps, if
unresolved, will lead to enhanced signal in an X-ray map and thus bias
the cluster temperature.  This will artificially increase the inferred
total mass.  However, the SZ data themselves are less sensitive to
clumping as the SZ signal is proportional to $n_{\mathrm{e}}$ rather
than $n_{\mathrm{e}}^2$.  Ultimately, the comparison of high
signal--to--noise SZ data with X-ray measurements will constrain the
level of clumping in clusters.


\subsection{Cluster Parameters}
\label{sec:params}

We discuss the constraints placed on core radius and
$\beta$--parameter by the VSA data, and also present results for the
gas mass, total mass and gas fractions calculated out to both
$r_{200}$ and $r_{500}$. For comparison, a summary of cluster
parameters derived from X-ray data is presented in Table
\ref{tab:x-params}.

\begin{table*}
\centering
\caption{Cluster parameters derived from X-ray data.  References are
[1] \protect\cite{Mason_predictions},
[2] \protect\cite{Mohr}, [3] \protect\cite{bham}} 

\setlength{\extrarowheight}{3pt}
\begin{tabular}{lcr@{}lcr@{}lcr@{}lr@{}l}
    \hline
	&\multicolumn{3}{c}{$r_{\rm{c}}$} &\multicolumn{4}{c}{$\beta$} 	&\multicolumn{4}{c}{$n_0$}\\ 
&\multicolumn{3}{c}{(arcmin)}&&&&&\multicolumn{2}{c}{($10^{-3}h_{100}^{1/2}\rm{cm}^{-3}$})&\multicolumn{2}{c}{($10^{-3}h_{50}^{1/2}\rm{cm}^{-3}$})\\    
	&[1]&&[3]		  	  &[1]&&[2]&[3]  		&[1]&&[2]&	\\
\hline
Coma	&$9.32\pm0.10$&&- 		&0.670&0.705&${}^{+0.046}_{-0.046}$	&-&4.51&${}^{+0.04}_{-0.04}$	&3.12&${}^{+0.04}_{-0.04}$\\
A1795	&$2.17\pm0.28$&4.01&${}^{+0.20}_{-0.21}$ &0.698	&0.790&${}^{+0.031}_{-0.032}$	&0.83$\pm{0.02}$&11.29&${}^{+0.61}_{-1.77}$ 	&29.9&${}^{+4.6}_{-1.5}$\\
A399	&$4.33\pm0.45$&1.89&${}^{+0.36}_{-0.36}$&0.742	&-&	&0.53$\pm{0.05}$&3.24&${}^{+0.14}_{-0.19}$ 	&-&\\
A401	&$2.26\pm0.41$&2.37&${}^{+0.09}_{-0.09}$&0.636	&0.606&${}^{+0.015}_{-0.016}$	&0.63$\pm{0.01}$&8.01&${}^{+0.56}_{-1.02}$	&5.87&${}^{+0.43}_{-0.27}$\\
A478	&$1.00\pm0.15$&2.34&${}^{+0.23}_{-0.23}$&0.638	&0.713&${}^{+0.030}_{-0.033}$	&0.75$\pm{0.01}$&28.9&${}^{+15.2}_{-3.9}$	&38.1&${}^{+3.3}_{-1.5}$\\
A2142	&$1.60\pm0.12$&3.14&${}^{+0.22}_{-0.22}$&0.635	&0.787&${}^{+0.082}_{-0.093}$	&0.74$\pm{0.01}$&15.03&${}^{+0.92}_{-1.07}$	&15.8&${}^{+1.7}_{-2.4}$\\
A2244	&$0.82\pm0.14$&&-			&0.580	&0.594&${}^{+0.061}_{-0.045}$	&- 		&17.73&${}^{+1.95}_{-2.65}$	&13.2&${}^{+1.9}_{-2.9}$\\
\hline
\end{tabular}
\setlength{\extrarowheight}{0pt}
\label{tab:x-params} 
\end{table*}

\begin{figure*}
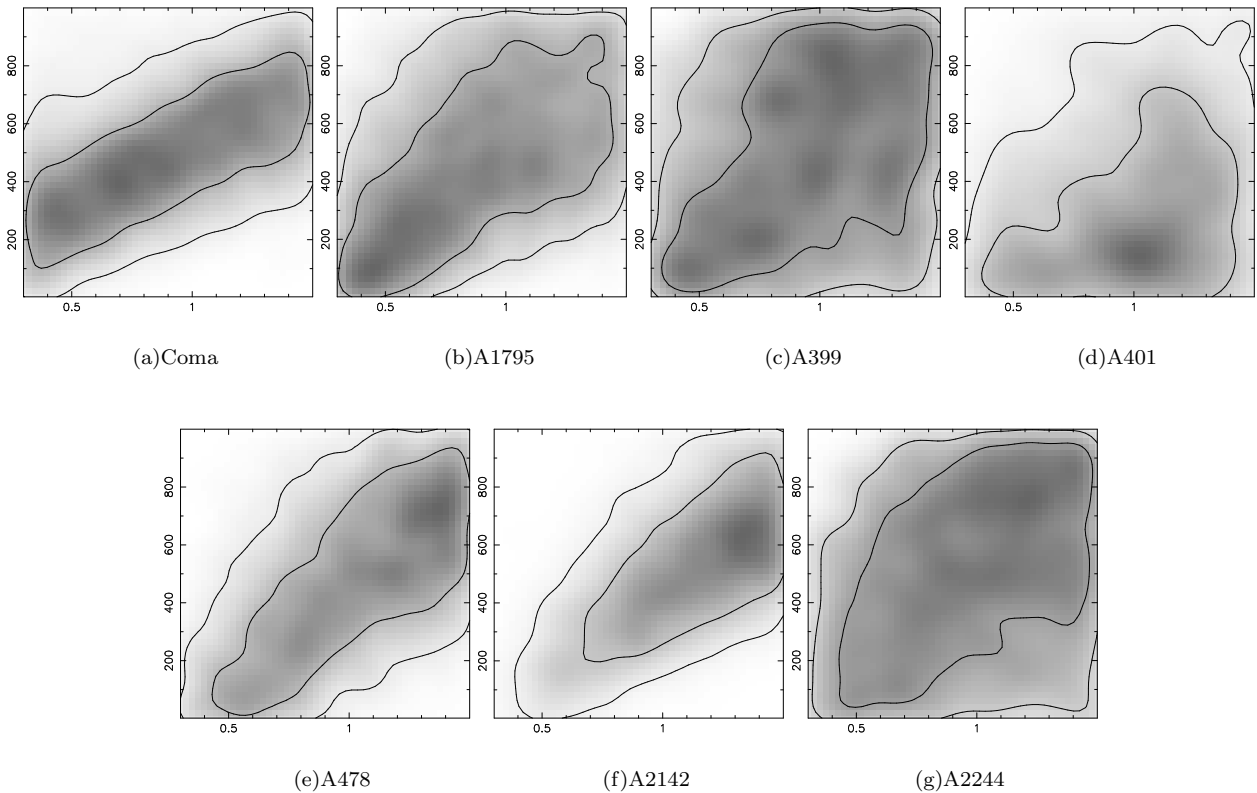

\begin{minipage}[b]{0.23\linewidth}
\begin{center}
\subfigure[Coma]{
\includegraphics*[width=\linewidth]{coma_b_r.ps}}
\end{center}
\end{minipage} 
\begin{minipage}[b]{0.23\linewidth}
\begin{center}
\subfigure[A1795]{
\includegraphics*[width=\linewidth]{a1795_b_r.ps}}
\end{center}
\end{minipage} 
\begin{minipage}[b]{0.23\linewidth}
\begin{center}
\subfigure[A399]{
\includegraphics*[width=\linewidth]{a399_b_r.ps}}
\end{center}
\end{minipage} 
\begin{minipage}[b]{0.23\linewidth}
\begin{center}
\subfigure[A401]{
\includegraphics*[width=\linewidth]{a401_b_r.ps}}
\end{center}
\end{minipage} 
\begin{minipage}[b]{0.23\linewidth}
\begin{center}
\subfigure[A478]{
\includegraphics*[width=\linewidth]{a478_b_r.ps}}
\end{center}
\end{minipage} 
\begin{minipage}[b]{0.23\linewidth}
\begin{center}
\subfigure[A2142]{
\includegraphics*[width=\linewidth]{a2142_b_r.ps}}
\end{center}
\end{minipage} 
\begin{minipage}[b]{0.23\linewidth}
\begin{center}
\subfigure[A2244]{
\includegraphics*[width=\linewidth]{a2244_b_r.ps}}
\end{center}
\end{minipage} 
\caption{Plots illustrating the constraints placed on $\beta$--parameter
and core radius by the cluster data.  In each plot, the x-axis is $\beta$
and the y-axis is core radius (kpc).  68\% and 90\% contours are shown.}
\label{fig:beta_rc}
\end{figure*}

We find, as anticipated, that the cluster parameters $\beta$ and
$r_{\rm{c}}$ are poorly constrained by the SZ data, as shown in Figure
\ref{fig:beta_rc}.  For Coma, A1795, A478 and A2142 there is
considerable degeneracy between the two parameters.  It is only
possible to place limits on the two parameters together -- little can be
said about them as separate entities. This is largely due to the
limited range of angular scales presented in this data, and indeed in
any SZ data to date.  Ideally, one would combine the VSA data with
observations on smaller angular scales.  This is impossible in this
case, as instruments such as the RT would completely resolve out
signal from the clusters in our sample.  AMI (see e.g. \cite{AMI})
will work over a larger range of angular scales and should start
to break this degeneracy.  A401, A399 and A2244 are not detected in
the cluster maps, so it is perhaps unsurprising that little constraint
can be placed upon the shape parameters by these data.

\begin{figure*}
\begin{minipage}[b]{1.0\linewidth}
\centering\epsfig{file=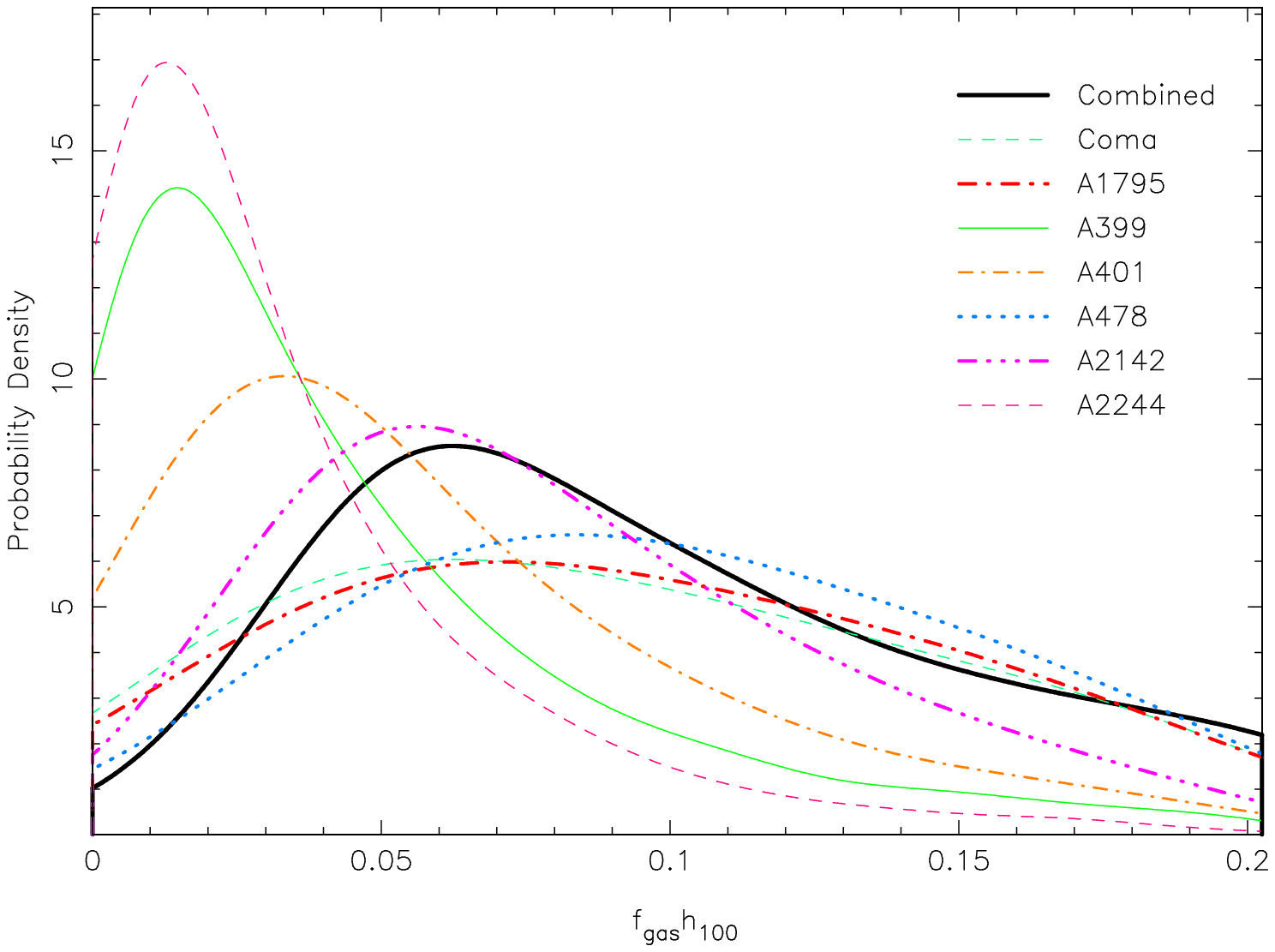,width=\linewidth,angle=0,clip=}
\end{minipage} 
\caption{Plot of the probability distributions for $f_{\rm{gas}}$ for
each cluster, and that derived from combining the full sample.}
\label{fig:f_gas}
\end{figure*}

We present the median of the probability distribution for the gas
mass, total mass and gas fraction for each cluster, evaluated to both
$r_{200}$ and $r_{500}$, in Table \ref{tab:fractions}.  The errors
quoted are the values of the $16.5^{\rm{th}}$ and the $83.5^{\rm{th}}$
percentiles.  We note that A1795, A478 and A2142 all favour a gas mass
of around $10^{14}M_{\sun}$. The Coma data allow very high gas masses.
This may be interpreted as the cluster position coinciding with a
negative feature in the CMB, thus making the SZ decrement appear
deeper.  The converse may be true for the other three clusters, in
that their SZ signals may be partially `obscured' by hot spots in the
CMB.  If this were true it would have the effect of reducing the
preferred values of the gas mass, and indeed these objects do allow
low values of this parameter.  (Note: although here we choose to
follow \cite{Myers_obs} in using X-ray temperatures from
\cite{Markevitch1998}, we recognise that more recent data are
available. Repeating the analysis using XMM-Newton temperatures
(\cite{Pointecouteau2004},
\cite{Sun2003}) we find that the resulting $f_{\mathrm{gas}}$ values 
are fully consistent with those presented in Figure \ref{fig:f_gas}
and Table \ref{tab:morefractions}.  Any variations are below the
random errors present in the VSA data).

\begin{figure*}
\begin{minipage}[b]{0.75\linewidth}
\mbox{
\hspace{-2.5cm}\epsfig{file=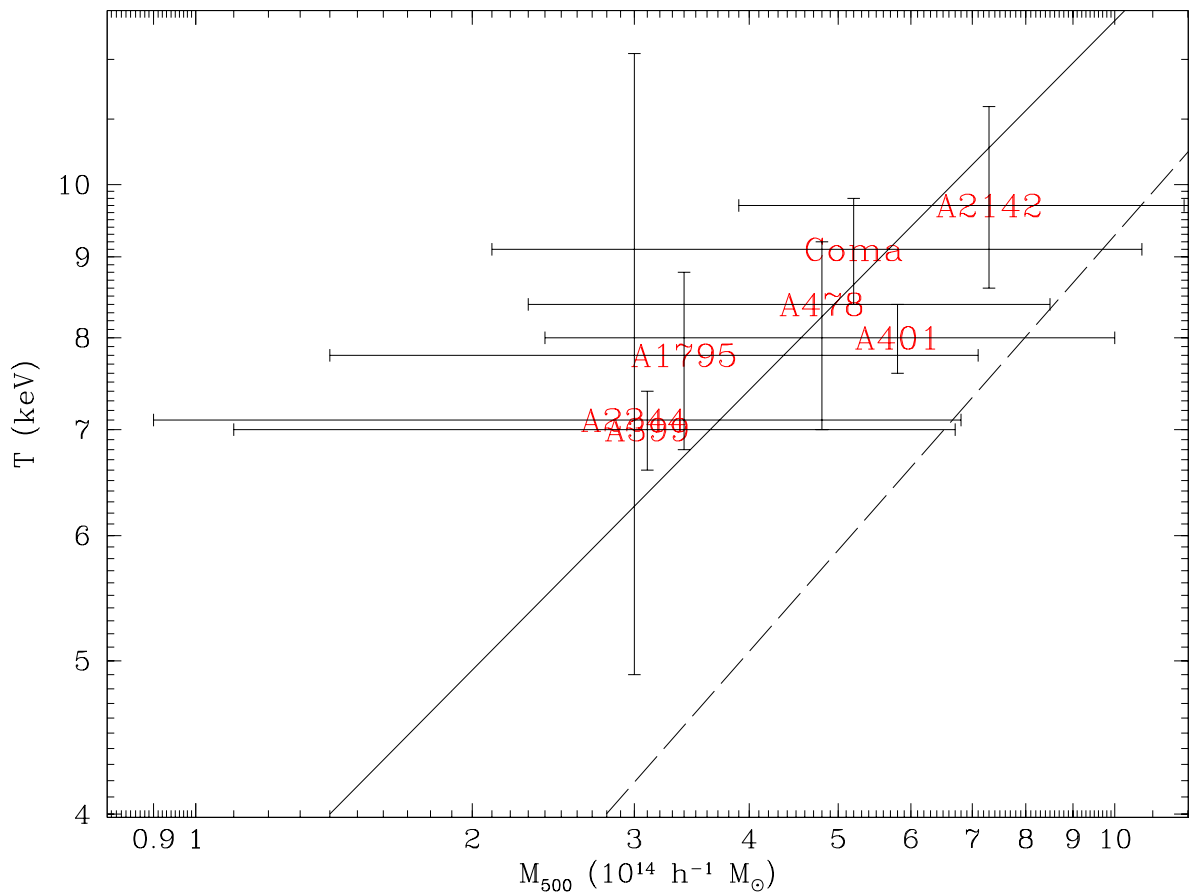, width=\linewidth, angle=-90, clip=}}
\end{minipage} 
\caption{The mass--temperature scaling relation derived from fitting
gas pressure profiles to the VSA SZ data.  The temperature shows the
X-ray temperatures given in Table \protect{\ref{tab:clusters}} and
also enters $M_{500}$ linearly.  The dashed line uses the
normalisation from hydrodynamical adiabatic simulations
(\protect\cite{evrard96}), and the solid line represents the best fit
$M-T$ relation of
\protect\cite{finoguenov01}.}
\label{fig:MT}
\end{figure*}

It is interesting to examine the constraints placed on the
relationship between total mass $M_{\rm{r}}$ and gas temperature by
the VSA SZ data.  In Figure \ref{fig:MT} we plot the X-ray determined
temperature and the total mass $M_{\rm{r}}$ derived using equation 6.
We expect, of course, some scatter on the values of $M_{\rm{r}}$ for
each cluster due to the CMB contamination of the SZ data.  After
examination of equation 6, we argue that the normalisation of our
$M-T$ relation is in fact mainly determined by the profile fitting
parameters $\beta$ and $r_{\rm{c}}$ derived from the VSA data, and
depends only weakly on $T_{\mathrm{X}}$ ($T_{\mathrm{X}}^{-1/2}$ for
the self--similar 3/2 slope of the M-T relation.)  This means in
Figure \ref{fig:MT} that the effect of any uncertainty in T (and
consequently in $M_{500}$) for a given set of $\beta$, $r_{\rm{c}}$
from the VSA will move the data points within their large error boxes
almost parallel to the slope of the $M-T$ relation.  For comparison we
plot the normalisation of the $M-T$ relations from hydrodynamical
adiabatic simulations (\cite{evrard96}) and X-ray cluster data
(\cite{finoguenov01}).  We calculate our normalisation constant for
$M\propto T^{3/2}$ to be $2.33^{+0.85}_{-0.78}\times 10^{13}$.  This
is in good agreement with the recent $M-T$ determinations derived from
X-ray data (\cite{X/ASF01}, \cite{Pratt}).  In a forthcoming paper we
intend to investigate the possibility of determining the $M-T$
relation from SZ without the use of an X-ray temperature.  Such an
$M-T$ relation, based on a measurement of the global gas pressure
distribution via the SZ effect, will be interesting to contrast with
X-ray measurements.  This kind of work will be very useful for the
interpretation of upcoming SZ cluster surveys.

The $f_{\rm{gas}}$ probability distributions are highly non-Gaussian,
and are plotted on the same axes in Figure \ref{fig:f_gas}.  The
errors quoted are the values of the $16.5^{\rm{th}}$ and the
$83.5^{\rm{th}}$ percentiles.  In order to compare values for
individual clusters, we summarise results from other experiments in
Table \ref{tab:morefractions}.  

\begin{table*}
\centering
\caption{Gas masses, total masses and gas fractions for the VSA
cluster sample evaluated to both $r_{200}$ and $r_{500}$.}

\setlength{\extrarowheight}{3pt}
\begin{tabular}{lrrrrr@{}lr@{}l}
    \hline
Cluster&$M_{\rm{gas}}(r_{200})h^2$	&$M_{\rm{gas}}(r_{500})h^2$
&$M_{\rm{r200}}h$		&$M_{\rm{r500}}h$
&\multicolumn{2}{c}{$f_{\rm{gas}}(r_{200})h$}
&\multicolumn{2}{c}{$f_{\rm{gas}}(r_{500})h$} \\
 	& $10^{13}M_{\sun}$	    &$10^{13}M_{\sun}$ 
	& $10^{14}M_{\sun}$	    &$10^{14}M_{\sun}$
	&& &&\\
\hline
Coma	&$15.4^{+9.0}_{-8.0}$	&$6.6^{+3.5}_{-3.0}$	
	&$10.9^{+10.0}_{-6.0}$	&$5.2^{+5.5}_{-3.1}$
&0.15&${}^{+ 0.28}_{- 0.10}$		&0.15&${}^{+ 0.17}_{- 0.09}$\\

A1795	&$8.5^{+3.9}_{-3.4}$	&$3.5^{+1.6}_{-1.7}$	
	&$7.9^{+6.5}_{-4.1}$	&$3.4^{+3.7}_{-2.0}$
&0.12&${}^{+0.15}_{-0.070}$		&0.11&${}^{+0.090}_{-0.060}$\\ 

A399	&$1.9^{+2.4}_{-1.3}$	&$0.7^{+1.1}_{-0.6}$	
	&$7.6^{+5.3}_{-3.9}$	&$3.1^{+3.6}_{-2.0}$
&0.030&${}^{+ 0.054}_{- 0.022}$		&0.028&${}^{+ 0.040}_{- 0.020}$\\

A401	&$5.0^{+3.0}_{-2.3}$	&$3.0^{+1.4}_{-1.4}$	
	&$10.7^{+6.0}_{-5.0}$	&$5.8^{+4.2}_{-3.4}$
&0.048&${}^{+ 0.074}_{- 0.028}$		&0.055&${}^{+ 0.055}_{- 0.029}$\\

A478	&$11.2^{+4.0}_{-4.0}$	&$5.7^{+2.1}_{-2.2}$	
	&$10.8^{+6.0}_{-5.0}$	&$4.8^{+3.7}_{-2.5}$
&0.12&${}^{+ 0.11}_{- 0.06}$		&0.13&${}^{+ 0.08}_{- 0.05}$\\

A2142	&$11.2^{+4.0}_{-3.0}$	&$6.1^{+1.7}_{-1.8}$	
	&$15.3^{+8.0}_{-6.0}$	&$7.3^{+4.6}_{-3.4}$
&0.074&${}^{+ 0.068}_{- 0.034}$		&0.086&${}^{+0.056}_{-0.035}$\\

A2244	&$ 1.3^{+1.6}_{-0.8}$	&$4.4^{+8.4}_{-3.7}$	
	&$7.5^{+5.8}_{-3.8}$	&$3.0^{+3.8}_{-2.1}$
&0.020&${}^{+ 0.039}_{- 0.015}$		&0.020&${}^{+0.031}_{-0.014}$\\
\hline
\end{tabular}
\setlength{\extrarowheight}{3pt}
\label{tab:fractions} 
\end{table*}

\begin{table*}
\centering
\caption{Gas fractions estimated within $R_0$ from SZ data ([1]
\protect\cite{Myers_obs}), and within $r_{500}$ from X-ray data ([2]
\protect\cite{Mason_predictions}, [3] \protect\cite{Mohr}, [4]
\protect\cite{fabian_fgas}).}

\setlength{\extrarowheight}{3pt}
\begin{tabular}{lccr@{}lcc}
    \hline

	&$f_{\rm{gas}}h$		&$R_0h$(Mpc)
		&\multicolumn{2}{c}{$f_{\rm{gas}}h^{3/2}$}
			&$f_{\rm{gas}}h^{3/2}_{50}$
				&$f_{\rm{gas}}h^{3/2}_{50}$\\
	&[1]				&[1]
		&\multicolumn{2}{c}{[2]}
			&[3]
				&[4]\\
\hline
Coma	&$0.063\pm0.017$&1.50	&0.0603&{}$\pm0.0028$		
			&0.177$\pm0.019$	&-\\
A1795	&-&-			&0.0477&{}$\pm0.0036$		
			&0.190$\pm0.008$	&$0.184\pm0.011$\\		
A399	&-&-			&0.0655&{}$\pm0.0032$		
						&-&-\\
A401	&-&-			&0.0794&{}$^{+0.0044}_{-0.0062}$ 
			&0.247$\pm0.012$	&$0.230\pm0.013$\\
A478	&$0.166\pm0.014$&0.976	&0.0760&{}$^{+0.0076}_{-0.0045}$ 
			&0.214$^{+0.012}_{-0.011}$ &$0.172\pm0.023$\\
A2142	&$0.060\pm0.011$&0.76	&0.0890&{}$^{+0.0064}_{-0.0091}$ 
			&0.227$^{+0.024}_{-0.017}$ &$0.255\pm0.033$\\
A2244	&-&-			&0.0739&{}$^{+0.0170}_{-0.0349}$ 
			&0.196$^{+0.061}_{-0.060}$ &$0.204\pm0.104$\\
\hline

\end{tabular}
\setlength{\extrarowheight}{0pt}
\label{tab:morefractions}
\end{table*}

We have combined the posterior probability density functions for each
cluster gas fraction as follows (see Marshall (2004), in preparation,
for more details).  Simulating the effect of simultaneously fitting
all our SZ data with the same global gas fraction $f_{\rm gas}$
requires dividing out the prior on the individual cluster gas fraction
(which can be derived from a set of MCMC samples with no data, see
\cite{anze}) and then multiplying the resulting effective
likelihoods together. Modulating this product by the prior on $f_{\rm
gas}$, which we take to be uniform over the range [0--0.2], gives us the
posterior pdf $\rm{P}$$(f_{\rm gas}|\rm{data})$. Moreover, keeping
track of the normalisations allows us to compute a relative
probability for the act of combination itself, that is, the ratio
$\rm{P}(\rm{data}|H^{\rm{global}})/\rm{P}(\rm{data}|H^{\rm{i}}))$,
where $\rm{H}^{\rm{i}}$ is the hypothesis `all clusters have independent
gas fractions~$f_{\rm{gas}}^{\rm{i}}$', whilst~$\rm{H}^{\rm{global}}$ is
the alternative hypothesis that `all clusters have the same gas
fraction~$f_{\rm gas}$'.

We first assume that \emph{all} our clusters have 
one true global gas fraction value, $f_{\rm{gas}}$ .  We combine the
individual probability density functions for all of our clusters,
including those with what would classically be called
non--detections.  We find
$f_{\rm{gas}}h_{100}=0.023^{+0.016}_{-0.012}$, with an evidence ratio in
favour of this all-encompassing combination of 
\begin{equation}
\frac{\rm{P}(\rm{data}|H^{\rm{global}})}
     {\rm{P}(\rm{data}|H^{\rm{i}})} = 4.4.
\label{result:A}
\end{equation}

We can also divide the data into two sets, those from detected
clusters and those from non--detections, and again investigate the
suitability of their combination.  Let
hypothesis~$\rm{H}^{\rm{global}}_{\rm{det}}$ consist of the assertions
that there is a global gas fraction $f_{\rm{gas}}$ exhibited by the
detected clusters, and that there is another gas fraction--like
parameter $X$ for the non--detections; we find the following evidence
ratios:
\begin{equation}
\frac{\rm{P}(\rm{data(detections)}|H^{\rm{global}}_{\rm{det}})}
     {\rm{P}(\rm{data(detections)}|H^{\rm{i}})} = 0.92,
\end{equation}
\begin{equation}
\frac{\rm{P}(\rm{data(non-detections)}|H^{\rm{global}}_{\rm{det}}) }
{\rm{P}(\rm{data(non-detections)}|H^{\rm{i}})} = 7.41.
\end{equation}
The former suggests that the data are not good enough to 
distinguish between the global gas fraction
hypothesis and that of all four detected 
clusters taking independent values
of~$f_{\rm{gas}}^{\rm{i}}$. However, the latter points strongly towards the 
combination of the non-detections' gas fractions. 
The overall evidence ratio from this `split sample' analysis is therefore:

\begin{align}
\frac{\rm{P}(\rm{data(all)}|H^{\rm{global}}_{\rm{det}})}
     {\rm{P}(\rm{data(all)}|H^{\rm{i}})}
&=
\frac{\rm{P}(\rm{data(detections)}|H^{\rm{global}}_{\rm{det}})}
     {\rm{P}(\rm{data(detections)}|H^{\rm{i}})} \notag\\
&\times
\frac{\rm{P}(\rm{data(non-detections)}|H^{\rm{global}}_{\rm{det}}) }
     {\rm{P}(\rm{data(non-detections)}|H^{\rm{i}})}  \notag\\
&=6.82
\label{result:B}
\end{align}

This is higher than the result in (\ref{result:A}), indicating that
the split sample analysis is more appropriate.  The interpretation is
that the detected clusters are telling us about a global cluster gas
fraction $f_{\rm{gas}}$, while the non--detections are telling us far
more about the primordial fluctuations (inappropriately parameterised
by~$X$).  Our `headline' result is therefore that from combining the
four detected clusters' gas fractions as above:
$f_{\rm{gas}}h_{100}=0.08^{+0.06}_{-0.04}$.  

In order to address the true value of a global $f_{\rm{gas}}$ we need
better data, which the likes of AMI (see e.g. \cite{AMI}), AMIBA (see
e.g. \cite{AMiBA}) and the SZA (see e.g. \cite{SZ/Moh++02}) should
provide.  We have, however, developed and demonstrated a useful method
for estimating the effect of, and for controlling, systematics. We could do
even better in estimating a universal $f_{\rm{gas}}$ if we were able
to use prior information (from X-rays and lensing) on the likely
detectability in SZ of each cluster.  This would require us to be able
to separate the `position' and `existence' implicit in the priors we
use; we are planning to attempt this.

We can also place formal constraints on $\Omega_{\rm{m}}h$ by assuming
that our estimation for $f_{\rm{gas}}h$ is indeed the global value.

\begin{equation}
f_{\rm{gas}}h = \frac {\Omega_{\rm{b}}h^2}{\Omega_{\rm{m}}h} 
\end{equation}

\noindent\cite{rafa} infer $\Omega_{\rm{b}}h^2$ and $h_{100}$ from VSA and
WMAP primordial CMB data, using a flat $\Lambda$CDM model.  We take
these values and find $\Omega_{\rm{m}}h=0.33^{+ 0.33}_{- 0.15}$.

Another implication concerns the clumping of the cluster gas. The
broad agreement here between $f_{\rm{gas}}$ values from X-ray and from
SZ, and as discussed in e.g. \cite{grego}, rules out significant
clumping.


\section{CONCLUSION}

We have investigated with the VSA Extended Array at $\approx34$\,GHz
the SZ effects towards seven nearby clusters that form a complete,
X-ray--flux--limited sample.

\begin{enumerate}
\item{Four of the clusters (Coma, A1795, A478, A2142) show SZ effects in
the map plane on scales of $\approx$20 arcmin of typically 6$\sigma$.}
\item{There is significant detection of CMB primordial structure at this
resolution, which is the likely cause of the three non-detections
(A399, A401, A2244).

We have analysed the data in the $uv$--plane, with X-ray priors on
positions and gas temperatures and radio priors on the sources, using
MCMC to estimate key cluster parameters in the context of a
$\beta$--model for the gas distribution. In this context, the CMB
primordial fluctuations are an additional source of Gaussian noise,
and are included in the model as a non--diagonal covariance matrix
derived from the known angular power spectrum.  We use the SZ data
(plus the priors) to give both the gas mass and, under the assumption
of hydrostatic equilibrium, the total mass. Although the data have
high random errors, the use of Bayesian methods, probability density
functions and marginalisation prevents bias in the results.}

\item{The degeneracy is evident between $\beta$ and core radius as
expected for such observations sensitive to SZ over a narrow
$\ell$--range. There are significant measurements of gas fractions in
the detected clusters.}

\item{We present a normalisation of the M-T relation derived from our
data which we find to be in good agreement with recent X-ray cluster
measurements.}

\item{Using the gas fraction probability density function for each
cluster, we have produced combined gas fractions for the four
detections, for the three non-detections, and for all seven. The
Bayesian evidence shows that the first is the correct one to use in
the context of trying to measure a low-$z$ global gas fraction. For
this we here find $f_{\rm{gas}} = 0.08^{+0.06}_{-0.04} h_{100}^{-1}$.}

\item{Gas fraction measurement by this SZ--based method is relatively
immune from the effect of primordial CMB anisotropy.  This is true
since the effect on gas mass tends to cancel the effect on total mass
on the narrow range of angular scale employed. Simulations show the
cancellation to be good for contaminant fluxes of $\pm50$\,mJy.}

\end{enumerate}

That the analysis method works as well as it does points the way towards
analysis of data from upcoming SZ telescopes.


\section{ACKNOWLEDGEMENTS}

The authors thank the referee for useful comments and suggestions. We
thank the staff of Jodrell Bank Observatory, the Mullard Radio
Astronomy Observatory and the Teide Observatory for assistance in the
day-to-day operation of the VSA.  We thank PPARC and the IAC for
funding and supporting the VSA project.  Partial financial support was
provided by the Spanish Ministry of Science and Technology, project
AYA2001-1657.  We thank Angela Taylor for processing and analysing the
source subtractor observations for Coma.  We also thank Clive
Dickinson, Will Grainger, Charlie McLachlan and Jonathan Zwart for
useful discussions. RK thanks Monique Arnaud and Sarah Church for
interesting conversations.  KL acknowledges support of a PPARC
studentship.


\bibliographystyle{mn2e}
\bibliography{references}


\bsp 

\label{lastpage}

\end{document}